\documentclass[onecolumn,aps,floatfix,pra,superscriptaddress,nofootinbib]{revtex4}
\usepackage{hyperref}
\usepackage{cleveref}
\usepackage{color}
\usepackage{graphicx}
\usepackage{bm}
\usepackage{amsmath}
\usepackage{enumerate}
\usepackage{upgreek}
\newcommand{\tr}{ \text{tr} }
\newcommand{\re}{ \text{ref} }
\newcommand{\erf}{ \text{erf} }
\newcommand{\erfc}{ \text{erfc} }

\newcommand{\cl}{ \text{cl} }
\newcommand{\qm}{ \text{q} }

\newcommand{\xd}{ \text{d} }
\newcommand{\ti}{ \tilde }
\newcommand{\ar}{ \text{A} }

\newcommand{\pa}{ \partial }
\newcommand{\hb}{ \hbar }
\newcommand{\si}{ \sigma }
\newcommand{\ga}{ \gamma }
\newcommand{\om}{ \omega }
\newcommand{\Om}{ \Omega }
\newcommand{\del}{ \delta }
\newcommand{\la}{ \langle }
\newcommand{\ra}{ \rangle }

\begin{document}
\title{

Stochastic Bohmian mechanics within the Schr\"{o}dinger-Langevin framework: A trajectory analysis of wave-packet dynamics in a fluctuative-dissipative medium



 

}
\author{S. V. Mousavi}
\email{vmousavi@qom.ac.ir}
\affiliation{Department of Physics, University of Qom, Ghadir Blvd., Qom 371614-6611, Iran}
\author{S. Miret-Art\'es}
\email{s.miret@iff.csic.es}
\affiliation{Instituto de F\'isica Fundamental, Consejo Superior de
Investigaciones Cient\'ificas, Serrano 123, 28006 Madrid, Spain}
\begin{abstract}
%
%
A Bohmian analysis of the so-called Schr\"{o}dinger-Langevin or Kostin nonlinear differential equation 
is provided to study how thermal fluctuations of the environment affects the dynamics of the wave packet from a 
quantum hydrodynamical point of view. In this way, after obtaining the Schr\"{o}dinger-Langevin-Bohm equation
from the Kostin equation its application to simple but physically insightful systems such as the Brownian-Bohmian motion, motion in a gravity field and transmission through a parabolic repeller is studied. 
If a time-dependent Gaussian ansatz for the probability density is assumed, the effect of thermal fluctuations  
together with thermal wave packets leads to Bohmian stochastic trajectories. From this trajectory based analysis, quantum and classical diffusion
coefficients for free particles, thermal arrival times for a linear potential and transmission probabilities and characteristic times such as arrival and dwell times for a parabolic repeller are then presented and discussed. 

\end{abstract}
\maketitle

%
\section{Introduction} \label{sec: intro}
%

It seems that the first attempt for generalizing Chandrasekhar's phenomenological theory \cite{Ch-RMP-1943} of Brownian motion to the quantum 
realm was made  by introducing a stochastic term in the Caldirola-Kanai framework \cite{Pa-JPA-1973}. It was shown that for the 
Boltzmann statistics, quantum stochastic dynamics lead to a diffusion constant which is the same as that obtained from the classical Brownian motion. 
The effect of noise on the Brownian motion of a quantum oscillator via the corresponding Caldirola-Kanai equation with a noise term was also studied 
\cite{Sv-TMF-1976}. 

On the other hand, Kostin \cite{Kostin-1972,Kostin-1975} derived heuristically the so-called non-linear Schr\"{o}dinger-Langevin equation 
(SL) from the quantum Langevin equation through Heisenberg position and momentum operators. This equation fulfills the unitarity condition, satisfies 
the uncertainty relation but violates the superposition principle due to its non-linear nature \cite{Razavy-2005}. Recently, a different generalized 
Schr\"{o}dinger equation has been proposed to describe dissipation in quantum systems \cite{Ch-EPJP-2017} where the 
Kostin equation without the random potential is a special case of this generalized equation.
Thanks to its straightforward formulation and its numerical simplicity, the Kostin equation ``can be considered as a solid candidate for 
effective description of open quantum systems hardly accessible to quantum master equations" \cite{KaGo-AOP-2016}.
Although this equation has been widely used in the literature to study pure dissipative effects 
\cite{ImKaGr-NPA-1975, KaGo-AOP-2016, MoMi-AOP-2018, Ch-IJQC-2018}, only a few works are found 
where the presence of a stochastic or random force is introduced \cite{KaGo-AOP-2016}. A generalized Schr\"odinger-Langevin
equation has been proposed in the literature for nonlinear interaction providing a state-dependent dissipation process exhibiting multiplicative noise
\cite{BaMi-AOP-2014}. Afterwards, this equation has been extended to a non-Markovian problem \cite{VaMoBa-AOP-2015}.

Bohmian mechanics \cite{Holland-book-1993,Sa-Mi-2012,Sa-Mi-2014} as an alternative interpretation of quantum mechanics has the 
advantage to give a clear picture of quantum phenomena in terms of trajectories in configuration space. This theory has many applications 
ranging from nano-scale systems to cosmology \cite{OrMo-book-2014}. Very recent applications to Maxwell's equations  in order to gain 
the independent manipulation of the optical phase evolution  and energy confinement \cite{YuPiPa-PRL-2018} as well as 
the tunnelling ionization and dynamics of the electron probability \cite{DoBa-PRA-2018}  have been studied. Its ability as a useful tool for 
open quantum systems has already been emphasized \cite{BASMO-EPJD-2014}.

In this work,  the Schr\"{o}dinger-Langevin-Bohm (SLB) equation is obtained by substituting the polar 
form of the wave function into the SL nonlinear differential equation \cite{NaMi-book-2017}. Then, by assuming a Gaussian ansatz as a solution for the SLB 
equation  for potentials of at most second order in space coordinate, the center of the wave packet is ruled by the {\it classical} Langevin equation 
and its width fulfills the generalized Pinney equation \cite{Pinney}. It is seen that the thermal fluctuations do not affect the width of the wave packet. 
The classical Langevin equation is solved by using a Gaussian delta-correlated white noise for the stochastic force which is typical for the 
Brownian motion. In Ref.  \cite{KaGo-AOP-2016}, quantum noise (white and colored) has been used to solve the Kostin equation for 
the simple harmonic oscillator. 
Once the SLB nonlinear differential equation is well established, the next step is to apply it to simple systems. The most basic problem to be
considered is the Brownian motion described by Bohmian stochastic trajectories. We have termed it the Brownian-Bohmian motion.
In the diffusion regime, time dependent diffusion coefficients issued from mean square displacements in the classical and Bohmian 
context are defined. At asymptotic times, both diffusion coefficients tend to be the same. This behavior can also be derived from the
velocity autocorrelation function \cite{Mi-2018}. Afterwards, thermal arrival times are also calculated for the falling particles under the presence 
of white noise. The last application considers the scattering of particles by a parabolic repeller. This potential function appears with different 
names in literature: inverted harmonic oscillator, repulsive harmonic oscillator or inverse oscillator. This type of barrier is paradigmatic in the sense 
that any interaction potential can be approximated by a parabolic function around its local maximum, rendering analytical solutions.
Exactly solvable models are interesting since they give general insights and provide an idea about accuracy of numerical results. The 
parabolic repeller is widely used in different contexts of physics ranging from condensed matter physics to cosmology. As it has been written 
in \cite{BeFe-AOP-2013} ``Although it started as an exercise from Landau’s book \cite{Lali-book-1958}, its physical applications have grown 
since the appearance of Barton’s Ph.D. Thesis (published in \cite{Ba-AOP-1986}), v.g., as an instability model, as a mapping of the 2D string 
theory \cite{YuKiCo-PS-2006}, or as a toy model to study early time evolution in inflationary models \cite{GuPi-PRD-1991}". Dissipative 
tunnelling through a parabolic repeller has been studied via quantum Langevin equation \cite{FoLeCo-PLA-1988}, realized by resistively 
shunted Josephson junctions, and using the Caldirola-Kanai \cite{Pa-JPA-1997} and Kostin \cite{MoMi-AOP-2018} approaches.
On the other hand, very recently, investigation of dissipative tunnelling times through different approaches have attracted many researchers 
\cite{KoNiTaHa-PLA-2007, BhRo-PRA-2012, BhRo-JMP-2013, KeLoEd-AOP-2017, Po-PRL-2017}. 
Transmission probabilities through tunnelling and above barrier as well as dwell times are then calculated at different bath temperatures.

This work is organized as follows. In Section II, the SLB nonlinear differential equation is introduced from the SL equation and a Gaussian ansatz
is proposed as a solution. An explicit expression for Bohmian stochastic trajectories are finally proposed.
Section III is devoted to non-dissipative and dissipative dynamics of thermal wave packets for simple potentials. Section
IV deals with the stochastic dynamics of thermal wave packets also for simple systems and special emphasis is on the so-called Brownian-Bohmian
motions where classical and quantum diffusion coefficients are defined through mean square displacements. In Section V, transmission probabilities
through a parabolic repeller and dwell times are studied for dissipative as well as stochastic dynamics. Finally, in Section VI some results and discussion
are presented.


%
%
\section{The Schr\"{o}dinger-Langevin-Bohm equation} \label{sec: Sch-La we}
%
%
In 1972, Kostin derived heuristically the so-called Schr\"{o}dinger-Langevin  equation which for one-dimensional problems reads
as \cite{Kostin-1972,Kostin-1975}
\begin{eqnarray} \label{eq: Sch_Lan}
i \hb \frac{\pa}{\pa t}\psi(x, t) &=& \left[ -\frac{\hb^2}{2m} \frac{\pa^2}{\pa x^2}
+ V(x, t) + V_r(x, t) + \frac{\gamma \hb }{2 i} \left( \ln \frac{\psi}{\psi^*} - \left \langle \ln \frac{\psi}{\psi^*} \right \rangle \right) 
\right] \psi(x, t)
\end{eqnarray}
to describe open quantum systems (in dissipative media with fluctuations) from the standard quantum Langevin equation. Here, $m$ is the 
particle mass which is subject to an external potential $V$ with $ \ga $ being the friction coefficient and $V_r$ a random potential linearly 
dependent on the particle position  and given by
\begin{eqnarray} \label{eq: random_pot}
V_r(x, t) &=& x~F_r(t) ,
\end{eqnarray}
$F_r(t)$ being a time-dependent random force. 
By substituting now the polar form of the wave function
\begin{eqnarray} \label{eq: tran_wf_polar}
\psi(x, t) &=& R(x, t) ~ e^{i S(x, t)/ \hb } 
\end{eqnarray}
into the Kostin equation (\ref{eq: Sch_Lan}), the resulting Schr\"{o}dinger-Langevin-Bohm (SLB) equation is expressed 
as \cite{NaMi-book-2017}
\begin{eqnarray} \label{kostin-bohmian} 
- \frac{\pa S}{\pa t} + i \hb \frac{1}{R} \frac{\pa R}{\pa t} 
& = & 
\frac{1}{2m} \left( \frac{\pa  S}{\pa x} \right)^2 + 
V(x, t) + V_r(x, t) + \gamma ( S - \langle S \rangle ) - \frac{\hb^2}{2m} \frac{1}{R} \frac{\pa^2 R}{\pa x^2}
\nonumber \\
&-& i \hb \frac{1}{2m} \left( \frac{\pa^2 S}{\pa x^2} + \frac{2}{R} \frac{\pa R}{\pa x} \frac{\pa S}{\pa x} \right) 
\end{eqnarray}
and by writing the real and imaginary parts separately, the following coupled equations are obtained,
\begin{eqnarray} \label{eq: con}
\frac{\pa  \rho}{\pa t} + \frac{ \pa }{\pa x} \left( \rho \frac{1}{m} \frac{\pa S}{\pa x} \right) &=&  0 , 
\end{eqnarray}
and
\begin{eqnarray}  \label{eq: HJ}
- \frac{\pa  S}{\pa t} &=& \frac{1}{2m} \left( \frac{\pa  S}{\pa x} \right)^2 + 
V(x, t) + V_r(x, t) + \gamma ( S - \langle S \rangle ) + Q (x, t)
\end{eqnarray}
which are the continuity and Hamilton-Jacobi equations, respectively, with  
\begin{eqnarray} 
\rho(x, t) &=& R^2(x, t)  \label{eq: rho}\\
Q(x, t) &=& - \frac{\hb^2}{2m} \frac{1}{R} \frac{\pa^2 R}{\pa x^2}  .
\end{eqnarray}
Here, $\rho$ is the probability density and $Q$ the so-called quantum potential. Now, if the velocity field is defined as
\begin{eqnarray} \label{eq: BM_vel}
v(x, t) &=& \frac{1}{m} \frac{\pa S}{\pa x} 
\end{eqnarray}
then, from the Hamilton-Jacobi equation, a Newtonian-like equation is easily deduced
\begin{eqnarray} \label{eq: Newton}
\frac{d v}{dt} &=& - \ga v - \frac{1}{m} \frac{\pa}{\pa x}( V + Q ) + \frac{F_r(t)}{m}
\end{eqnarray}
where the relation
\begin{eqnarray} 
\frac{d}{dt} &=& \frac{\pa}{\pa t} + v \frac{\pa}{\pa x}
\end{eqnarray}
has been used.

%
%
%

By imposing now a time-dependent Gaussian ansatz for the probability density 
\begin{eqnarray} \label{eq: rho_ansatz}
\rho (x, t) &=& \frac{1}{\sqrt{2\pi} ~ \si(t)} \exp \left[ -\frac{(x-q(t))^2}{2 \si^2(t)} \right] ,
\end{eqnarray}
where $q(t) = \int dx ~ x \rho(x, t) $ is the time dependent expectation value of the position operator which follows the center of the 
Gaussian wave packet and $\si(t)$ gives its width, Eq. (\ref{eq: rho_ansatz}) satisfies the continuity equation (\ref{eq: con}) with the 
velocity field
\begin{eqnarray} \label{eq: vel}
v(x, t) &=& \dot{q}(t) + \frac{ \dot{\si}(t) }{ \si(t) } (x-q(t))
\end{eqnarray}
We assume now that at each space point at each instant, this velocity field defines the tangent to a possible particle trajectory passing through that point, $ \dot{x} = v(x, t) |_{x = x(x^{(0)}, t)}$. Then, from this relation and Eq.(\ref{eq: vel}) Bohmian stochastic trajectories are obtained as
\begin{eqnarray} \label{eq: BM_traj}
x(x^{(0)}, t) &=& q(t) + \frac{\si(t)}{\si(0)}( x^{(0)} - q(0) ) ,
\end{eqnarray}
where $q(0)$ and $x^{(0)}$ are the initial values of $q(t)$ and position of the particle, respectively.

By introducing Eqs. (\ref{eq: rho_ansatz}) and (\ref{eq: vel}) into Eq. (\ref{eq: Newton}) one obtains
\begin{eqnarray}
\left( \ddot{ \si }(t) + \ga \dot{ \si }(t) - \frac{ \hb^2 }{ 4m^2 \si(t)^3} \right) (x-q(t)) + \si(t) ( \ddot{q}(t) + \ga \dot{q}(t))
&=& \si(t) \left( - \frac{1}{m} \frac{\pa V}{\pa x} + \frac{F_r(t)}{m} \right)
\end{eqnarray}
Now by using the wave packet approximation i.e., assuming an enough slowly varying potential in the range of the wave packet, the interaction potential is Taylor expanded around the classical trajectory $q(t)$ up to second order.  Then by using the condition for linear independence of different powers 
of $ (x - q(t)) $ and after some quite straightforward algebra we reach the following uncoupled equations for $q(t)$ and $\si(t)$
\begin{eqnarray} 
\ddot{q}(t) + \ga \dot{q}(t) + \frac{1}{m} \frac{\pa V}{\pa x}\bigg|_{x=q(t)} &=& \frac{F_r(t)}{m}  , 
\label{eq: xt} 
\\
\ddot{ \si }(t) + \ga \dot{ \si }(t) - \frac{ \hb^2 }{ 4m^2 \si(t)^3} +  \frac{\si(t)}{m} \frac{\pa^2 V}{\pa x^2}\bigg|_{x=q(t)} &=& 0 . 
\label{eq: delt}
\end{eqnarray}
Apparently, this approximation is exact for potentials of at most quadratic order in space coordinate. 
Eq. (\ref{eq: xt}) is the well-known classical Langevin equation where 
the fluctuating force or noise term is responsible for the numerous microscopic impacts undergone by the system due to the incessant 
scattering of environmental particles (thermal bath). $F_r(t) $ is generally taken to be a Gaussian random process (classical white noise) and is 
usually characterized by the following two properties
\begin{eqnarray} 
\la F_r(t) \ra &=& 0 , \label{eq: white} \\
\la F_r(0) F_r(t) \ra &=& 2m \ga k_B T \del(t)  \label{eq: fluc-diss}
\end{eqnarray}
with $k_B$ being the Boltzmann constant, $T$ the temperature of the bath and $ \del(t) $ the Dirac delta function. This relation comes from 
the fluctuation-dissipation theorem, which relates the fluctuating and dissipation processes in order to obtain a balance between them. 
This balance is correct if the equilibrium distribution of the system is Boltzmannian and the temperature reached by the system, $T_s$, 
equals the bath temperature $T$  \cite{KaGo-AOP-2016}. Furthermore, Eq. (\ref{eq: delt}) is the so-called generalized Pinney equation 
\cite{Pinney, NaMi-book-2017}. The Pinney equation is also known as Ermakov equation and appears, for example, in the
process of cooling down atoms in a harmonic trap \cite{Mu-PRL}.
It is clearly seen that noise just affects the motion of the center of the wave packet, not of its width.
It should also be noted that Bohmian mechanics provides the same results as standard quantum theory if the initial positions $ x^{(0)} $ 
distribute according to the Born rule. In this context, the initial velocity of the particle is specified by the guidance condition
$ v^{(0)} = \frac{1}{m}\frac{\pa}{\pa x} S(x^{(0)}, 0) $ which from Eq. (\ref{eq: vel}) and using the initial condition 
$ \dot{\si}(0) = 0 $, it reduces to $ v^{(0)} = \dot{q}(0) $.

In the following, dissipative and stochastic dynamics will be analyzed and discussed for some simple unbound systems
and thus we will assume a Maxwell-Boltzmann distribution for the initial velocities for the scattering particles.

\section{Non-dissipative and dissipative dynamics of thermal wave packets} \label{sec: thermal_wave}

Thermal wave packets are going to be considered for an ensemble of non-interacting particles from a thermal distribution of initial velocities. 
The initial state is taken to be a mixed ensemble of Gaussian wave packets with weights given by the Maxwell-Boltzmann distribution. 
Throughout this section, to avoid confusions, the temperature of the system will be denoted by $T_s$.
%
We study first the non-dissipative case and then take into account dissipation by means of the SLB equation.

\subsection{Free friction} \label{sec: non-dis}

Consider an ensemble of noninteracting particles where each particle is initially described by the state $ | \psi_{\dot{q}(0)}(0) \rangle $ 
with $\dot{q}(0)$ being the central velocity of the corresponding wave packet 
\begin{eqnarray} \label{eq: psi_gauss_0}
\psi_{\dot{q}(0)}(x, 0) &=& \frac{1}{(2\pi \si_0^2)^{1/4}} \exp \left[ -\frac{(x-q(0))^2}{4 \si_0^2} + i \frac{m \dot{q}(0)}{\hb} x \right] .
\end{eqnarray}
Particles are assumed to have a Maxwell-Boltzmann distribution of initial velocities given by
\begin{eqnarray} \label{eq: MB_dis}
f_{T_s}(\dot{q}(0)) &=& \sqrt{ \frac{m}{2\pi k_B T_s} } \exp \left[ -\frac{m \dot{q}(0)^2}{2 k_B T_s} \right]    ,
\end{eqnarray}
with the same mass, $m$.
Our mixed ensemble is described by \cite{Sakurai-book-1999}
\begin{eqnarray} \label{eq: den_mat_0}
\hat{\rho}_{T_s}(0) &=& \int_{-\infty}^{\infty} d\dot{q}(0) ~ f_{T_s}(\dot{q}(0)) | \psi_{\dot{q}(0)}(0) \rangle \langle \psi_{\dot{q}(0)}(0) | 
\end{eqnarray}
whose time evolution is given by the von Neumann equation of motion 
\begin{eqnarray} \label{eq: voNewmann}
i \hb \frac{\pa \hat{\rho}}{\pa t} &=& [\hat{H}, \hat{\rho}]
\end{eqnarray}
and expressed as
\begin{eqnarray} \label{eq: den_mat_t}
\hat{\rho}_{T_s}(t) &=& e^{-i \hat{H} t / \hb } \hat{\rho}_{T_s}(0) e^{i \hat{H} t / \hb } =
\int_{-\infty}^{\infty} d\dot{q}(0) ~ f_{T_s}(\dot{q}(0)) ~ | \psi_{\dot{q}(0)}(t) \rangle \langle \psi_{\dot{q}(0)}(t) |
\end{eqnarray} 
where $| \psi_{\dot{q}(0)}(t) \rangle = e^{-i \hat{H} t / \hb } | \psi_{\dot{q}(0)}(0) \rangle$, $ \hat{H} $ being the Hamiltonian of the system.
%
%
%
The matrix elements of the thermal density operator (\ref{eq: den_mat_t}) in the coordinate representation are given by 
\begin{eqnarray} \label{eq: mat_elem_t}
\rho_{T_s}(x, x', t) &=&  
\int_{-\infty}^{\infty} d\dot{q}(0) ~ f_{T_s}(\dot{q}(0)) ~ \psi_{\dot{q}(0)}(x ,t) \psi^*_{\dot{q}(0)}(x', t) .
\end{eqnarray}
In particular,  the diagonal elements of the density matrix, which are interpreted as a probability distribution, have the Gaussian form
\begin{eqnarray} \label{eq: diag_mat_elem_t}
\rho_{T_s}(x, t) &=&  \rho_{T_s}(x, x', t) \bigg|_{x'=x}
= \frac{1}{\sqrt{2\pi} \si_{T_s}(t)} \exp\left[{-\frac{( x - X(t) )^2}{2 \si_{T_s}(t)^2}} \right]
\end{eqnarray}
where the center of the wave packet follows the thermal-averaged trajectory $ X(t) = \la q(t) \ra $, $ q(t) $ being again the center of the wave 
packet $ |\psi_{\dot{q}(0)}(x, t)|^2 $;  its width has now a temperature contribution.
For the expectation value of any observable $\hat{A} = \hat{A}(\hat{x}, \hat{p})$, we have that 
\begin{eqnarray} \label{eq: thermal_expect_value}
\langle \hat{A} \rangle_{T_s}(t) &=&
\text{Tr}(\hat{\rho}_{T_s}(t) \hat{A})
= \int_{-\infty}^{\infty} d\dot{q}(0) ~ f_{T_s}(\dot{q}(0)) 
\int_{-\infty}^{\infty} dx ~ \psi^*_{\dot{q}(0)}(x, t) ~ A\left(x, -i\hb \frac{\pa}{\pa x} \right) ~  \psi_{\dot{q}(0)}(x ,t) .
\end{eqnarray}
As a special case, the expectation value of the space coordinate $\hat{x}$ is 
\begin{eqnarray} \label{eq: thermal_x_expect_value}
\langle \hat{x} \rangle_{T_s}(t) &=&  
\int_{-\infty}^{\infty} d\dot{q}(0) ~ f_{T_s}(\dot{q}(0)) \int_{-\infty}^{\infty} dx ~ \psi^*_{\dot{q}(0)}(x, t) ~ x ~  \psi_{\dot{q}(0)}(x ,t) \nonumber \\
&=&
\int_{-\infty}^{\infty} dx ~ x ~ \rho_{T_s}(x, t)   .
\end{eqnarray}
This relation confirms the interpretation of the diagonal matrix elements of the density operator as the probability density.

\subsubsection{Free particles}

The propagator for the free particle with Hamiltonian $\hat{H} = \frac{\hat{p}^2}{2m}$ is given by
\begin{eqnarray} \label{eq: free_prop}
\langle x | e^{-i \hat{H} t / \hb } | x' \rangle &=&  
\sqrt{ \frac{m}{2\pi i \hb t} } \exp \left[ \frac{i m}{ 2 \hb t} (x-x')^2 \right] 
\end{eqnarray}
and the wave function with initial velocity $\dot{q}(0)$ is then written as
\begin{eqnarray} \label{eq: wave_t_free}
\psi_{\dot{q}(0)}(x, t) &=& \int_{-\infty}^{\infty} dx' ~ \langle x | e^{-i \hat{H} t / \hb } | x' \rangle ~
\psi_{\dot{q}(0)}(x', 0)  
\nonumber \\
&=& \frac{1}{(2\pi)^{1/4} \sqrt{s_t}}
\exp \left[ \frac{i m}{2 \hb t} \left( x^2 + \frac{i \hbar t}{2 m \si_0^2} q(0)^2 \right) - \frac{\si_0}{s_t} \left( x-q(0)-\dot{q}(0) t + \frac{s_t}{\si_0} q(0) \right)^2 \right]
\end{eqnarray}
where the complex width is
\begin{eqnarray} \label{eq: st}
s_t &=& \si_0 \left( 1 + i \frac{\hb t}{2 m \si_0^2} \right) .
\end{eqnarray}
From Eq. (\ref{eq: wave_t_free}), the thermal Gaussian shape (\ref{eq: diag_mat_elem_t}) for the diagonal elements of density matrix is 
easily obtained from
\begin{eqnarray} 
X(t) &=& q(0) \label{eq: xt_free} \\
\si_{T_s}(t) &=& \si_0 \sqrt{ 1 + \left( \frac{ \hb^2 }{4 m^2 \si_0^4} + \frac{k_B T_s}{m\si_0^2} \right)t^2 } .
\label{eq: sigmat_free}
\end{eqnarray}
As one clearly sees there is a temperature-dependence contribution to the width. For a given time, the width increases with temperature. 
Now, from Eq. (\ref{eq: thermal_expect_value}), the first two moments of the momentum distribution for a given temperature are
\begin{eqnarray} 
\langle \hat{p} \rangle_{T_s}(t) &=& 0 \label{eq: p_ex_free} \\
\langle \hat{p}^2 \rangle_{T_s}(t) &=& \frac{\hb^2}{4\si_0^2} + m k_B T_s \label{eq: p2_ex_free}
\end{eqnarray}
and the corresponding uncertainty is then given by
\begin{eqnarray} \label{eq: p_uncer_free}
\Sigma_{T_s} &=& \sqrt{ \langle \hat{p}^2 \rangle_{T_s}(t) - \langle \hat{p} \rangle^2_{T_s}(t) } = \sqrt{ \frac{\hb^2}{4\si_0^2} + m k_B T_s } 
\end{eqnarray}
which is also dependent on the temperature, but not on time.

As a consistency check, the thermal Wigner distribution function which is defined by
\begin{eqnarray} \label{eq: wigner}
W_{T_s}(x, p, t) &=& \frac{1}{\pi \hb } \int_{-\infty}^{\infty} dy~ 
\langle x+y | \hat{\rho}_{T_s}(t) | x-y \rangle
e^{i 2 p y/ \hb } \nonumber \\
&=&
\frac{1}{\pi \hb } \int_{-\infty}^{\infty} dv_0 ~ f_{T_s}(\dot{q}(0)) \int_{-\infty}^{\infty} dy ~
\psi_{\dot{q}(0)}(x+y, t) ~ \psi^*_{\dot{q}(0)}(x-y, t) e^{i 2 p y/ \hb } 
\end{eqnarray}
can be calculated for the free case reaching
\begin{eqnarray} \label{eq: W_free}
W_{T_s}(x, p, t) &=& \frac{1}{ \sqrt{ \pi(\hb^2 + 4 m \si_0^2 k_B T_s) } } 
\exp \left[ - \frac{2 \si_0^2 p^2}{ \hb^2 + 4 m \si_0^2 k_B T_s } - 
\frac{ ( m(x-q(0)) + p t )^2 }{ 2m^2 \si_0^2 }
 \right].
\end{eqnarray}
By integrating over the spatial coordinate $x$, it leads to the momentum distribution
\begin{eqnarray} \label{eq: mom_dis_free}
\Pi_{T_s}(p) &=& \frac{1}{ \sqrt{2\pi} \Sigma_{T_s} } \exp \left[ - \frac{ p^2 }{ 2 \Sigma_{T_s}^2 } \right]
\end{eqnarray}
with $\Sigma_{T_s}$ given by Eq. (\ref{eq: p_uncer_free}).

\subsubsection{Linear potential}

The propagator for a linear potential with Hamiltonian $\hat{H} = \frac{\hat{p}^2}{2m} + K x $ is
\begin{eqnarray} \label{eq: lin_prop}
\langle x | e^{-i \hat{H} t / \hb } | x' \rangle &=&  
\sqrt{ \frac{m}{2\pi i \hb t} } \exp \left[ \frac{i m}{ 2 \hb t} (x-x')^2 - i \frac{K t}{2\hb}(x+x')   
- i \frac{K^2}{24 m \hb} t^3 \right]    .
\end{eqnarray}
From this expression, one obtains the Gaussian function (\ref{eq: diag_mat_elem_t}) with a width given by Eq. (\ref{eq: sigmat_free}) and the center 
of the thermal wave packet follows the classical trajectory

\begin{eqnarray} \label{eq: xt_linear}
X(t) &=& q(0) - \frac{K t^2}{2 m}    .
\end{eqnarray}
Again, the first two moments of the momentum distribution are 
\begin{eqnarray} 
\langle \hat{p} \rangle_{T_s}(t) &=& - K t \label{eq: p_ex_linear} \\
\langle \hat{p}^2 \rangle_{T_s}(t) &=& \frac{\hb^2}{4\si_0^2} + m k_B T_s + K^2 t^2 \label{eq: p2_ex_linear}
\end{eqnarray}
and thus, the corresponding uncertainty takes again the form of Eq. (\ref{eq: p_uncer_free}).

\subsubsection{Parabolic repeller potential}

The propagator for the inverted harmonic oscillator  with Hamiltonian $\hat{H} = \frac{\hat{p}^2}{2m} -\frac{1}{2} m \om^2 x^2 $ is
\begin{eqnarray} \label{eq: para_prop}
\langle x | e^{-i \hat{H} t / \hb } | x' \rangle &=&  
\sqrt{ \frac{ m \om }{2\pi i \hb \sinh \om t} } \exp \left[ \frac{i m \om}{ 2 \hb \sinh \om t} ( (x-x')^2 \cosh \om t - 2 x x' ) \right] 
\end{eqnarray}
and the center of the  thermal Gaussian function (\ref{eq: diag_mat_elem_t}) and its width are in this case
\begin{eqnarray} 
X(t) &=& q(0) \cosh( \om t ) \label{eq: xt_para} \\
\si_{T_s}(t) &=& \si_0 \sqrt{
\cosh^2( \om t ) + \left(  \frac{ \hb^2 }{ 4 m^2 \si_0^4 } + \frac{k_B T_s}{m \si_0^2}\right) \frac{ \sinh^2( \om t ) }{ \om^2 }  .
}\label{eq: sigmat_para}
\end{eqnarray}
Again, from the first two moments of the momentum distribution
\begin{eqnarray} 
\langle \hat{p} \rangle_{T_s}(t) &=& m \om q(0) \sinh( \om t )  \label{eq: p_ex_para} \\
\langle \hat{p}^2 \rangle_{T_s}(t) &=& \left( \frac{\hb^2}{4\si_0^2} + m k_B T_s \right) \cosh^2( \om t ) 
+ m^2 (q(0)^2 + \si_0^2) \om^2 \sinh^2( \om t )
\label{eq: p2_ex_para}
\end{eqnarray}
the uncertainty is written as
\begin{eqnarray} \label{eq: p_uncer_para}
\Sigma_{T_s}(t) &=& \sqrt{ \left( \frac{\hb^2}{4\si_0^2} + m k_B T_s \right) \cosh^2( \om t ) 
+ m^2 \si_0^2 \om^2 \sinh^2( \om t )
} 
\end{eqnarray}
which is now time and temperature dependent.

 \subsection{Influence of dissipation} \label{sec: dissipation}

In the absence of the fluctuating force, the Kostin equation (\ref{eq: Sch_Lan}) is rewritten as
\begin{eqnarray}  \label{eq: Kostin}
i \hbar \frac{\partial}{\partial t}\psi_{\dot{q}(0), \ga}(x, t) &=& \left[ -\frac{\hbar^2}{2m} \frac{\partial^2}{\partial x^2}
+ V(x) + \frac{\gamma \hbar}{2 i} \left( \ln \frac{\psi_{\dot{q}(0), \ga}}{\psi_{\dot{q}(0), \ga}^*} - \left \langle \ln \frac{\psi_{\dot{q}(0), \ga}}
{\psi_{\dot{q}(0), \ga}^*} \right \rangle \right) 
\right] \psi_{\dot{q}(0), \ga}(x, t) .
\end{eqnarray}
As it was mentioned above, for the quadratic potential
\begin{eqnarray} \label{eq: quad_pot}
V(x) &=& m g x - \frac{1}{2} m \om^2 x^2
\end{eqnarray}
the {\it exact} Gaussian solution (\ref{eq: rho_ansatz}) is now expressed as
\begin{eqnarray} \label{eq: Gauss_an}
 |\psi_{\dot{q}(0), \ga}(x, t)|^2 &=& \frac{1}{\sqrt{2\pi} \si_{\ga}(t)} \exp \left[ -\frac{(x-q_{\ga}(t))^2}{2 \si_{\ga}(t)^2} \right] ,
\end{eqnarray}
where the center of the wave packet, which is the solution of Eq. (\ref{eq: xt}) without the fluctuating force, is expressed by
\begin{eqnarray} \label{eq: xbar}
q_{\ga}(t) &=& - \frac{ g }{ \om^2 } + \left( q(0) + \frac{ g }{ \om^2 } \right) \left[ \cosh \Om t + \frac{\ga}{2} \frac{\sinh \Om t}{\Om} \right] e^{-\ga t /2} + \dot{q}(0) ~ \frac{\sinh \Om t}{\Om}~e^{-\ga t /2},
\end{eqnarray}
with the frequency $\Om$ given by
\begin{eqnarray} \label{eq: Omega}
\Om &=& \sqrt{\om^2 + \ga^2/4} .
\end{eqnarray}
The width is the solution of the generalized Pinney equation (\ref{eq: delt}),
\begin{eqnarray} \label{eq: Pinney}
\ddot{ \si }_{\ga}(t) + \ga \dot{ \si }_{\ga}(t) - \frac{\hb^2}{4 m^2  \si_{\ga}(t)^3} - \om^2  \si_{\ga}(t)  &=& 0  
\end{eqnarray}
which has no analytical solution.

Thus, in a dissipative medium, the time evolution of the state (\ref{eq: den_mat_0}) yields  
\begin{eqnarray} \label{eq: prob_den_gamma}
\rho_{\ga, T_s}(x, t) &=& \la x | \rho_{\ga, T_s}(t) | x \ra
= \frac{ 1 }{ \sqrt{2\pi} \si_{\ga, T_s}(t) } 
\exp \left[ -\frac{( x- X_{\ga}(t) )^2}{2 \si_{\ga, T_s}(t)^2} \right] ,
\end{eqnarray}
for the thermal probability density, where
\begin{eqnarray} 
X_{\ga}(t) &=&  - \frac{ g }{ \om^2 } + \left( q(0) + \frac{ g }{ \om^2 } \right) \left[ \cosh \Om t + \frac{\ga}{2} \frac{\sinh \Om t}{\Om} \right] e^{-\ga t /2}  \label{eq: Xt_gamma}
\\
\si_{\ga, T_s}(t) &=& \sqrt{ \si_{\ga}(t)^2 + e^{-\ga t} ~ \frac{k_B T_s}{m \Om^2} \sinh^2(\Om t)} \label{eq: var}     .
\end{eqnarray}
%

%
\section{Stochastic dynamics of thermal wave packets} \label{sec: LE}
%

When the correct balance between the fluctuating and dissipation processes occurred then $T_s=T$, and knowing that 
the Gaussian solution of the SLB equation is exact for quadratic potentials (\ref{eq: quad_pot}), Eq. (\ref{eq: xt}) takes now the form
\begin{eqnarray} \label{eq: xt_general} 
\ddot{q}(t) + \ga \dot{q}(t) + g - \om^2 q(t)  &=& \frac{F_r(t)}{m} , 
\end{eqnarray}
where $ g $ and $ \om $ are constants with dimensions of acceleration and frequency, respectively.
The solution of the equation (\ref{eq: xt_general}) is
\begin{eqnarray}  \label{eq: xt_gen_sol} 
q(t) &=& q_{\ga}(t) + \frac{1}{m \Om} \int_0^t d\tau F_r(\tau) \sinh \Om (t-\tau)  e^{-\ga (t-\tau) /2}.
\end{eqnarray}
where $ q_{\ga}(t) $ is given by Eq. (\ref{eq: xbar}). The velocity of the wave packet center can also be written as
\begin{eqnarray} \label{eq: vt_gen_sol}
\dot{q}(t) &=& \left( q(0) - \frac{g}{\om^2} \right) \frac{\om^2}{\Om} \sinh (\Om t)  e^{-\ga t /2}
+ \dot{q}(0) \left[ \cosh (\Om t) - \frac{\ga}{2 \Om} \sinh (\Om t) \right] e^{-\ga t /2}
\\
&+& \frac{1}{m}  \int_0^t d\tau F_r(\tau) e^{-\ga (t-\tau) /2} 
\left\{\cosh (\Om (t-\tau)) - \frac{\ga}{2 \Om} \sinh (\Om (t-\tau))   \right\}   .
\end{eqnarray}
When $ F_r = 0 $, Eqs. (\ref{eq: xt_gen_sol}) and (\ref{eq: vt_gen_sol}) are the solution of purely (dissipative) deterministic equations of 
motion. Thus, without loss of generality, they can be expressed as
\begin{eqnarray} 
q(t) &=& q_d(t) + q_s(t) \label{eq: xt_split}
\\
\dot{q}(t) &=& \dot{q}_d(t) + \dot{q}_s(t) \label{eq: vt_split}
\end{eqnarray}
where the sub-index $d$ stands for deterministic and $s$ for stochastic. Quantities of interest are average values. 
By using the property (\ref{eq: white}) of the stochastic force we have that
\begin{eqnarray} 
\la q(t) \ra &=& q_d(t)  
\\
\la \dot{q}(t) \ra &=& \dot{q}_d(t)   
\\
\la q^2(t) \ra &=& q_d^2(t) + \la q_s^2(t) \ra
\\
\la \dot{q}^2(t) \ra &=& \dot{q}_d^2(t) + \la \dot{q}_s^2(t) \ra
\end{eqnarray}
where property (\ref{eq: fluc-diss}) is used to compute average values of squares. In the following, we separately consider three cases:
a force-free field, gravitational field and  parabolic repeller potential. 

%
\subsection{Free particle. The Brownian-Bohmian motion} \label{subsec: free}
%

For a free particle, $ V(x) = 0 $, and from Eq. (\ref{eq: xt_gen_sol}), the solution to the equation of motion is given by
\begin{eqnarray} \label{eq: xt_free_formal_sol} 
q(t) &=& q(0) + \dot{q}(0) \frac{1}{\ga} (1- e^{-\ga t} ) 
+ \frac{1}{m \ga} \int_0^t d\tau F_r(\tau) e^{-\ga ( t - \tau) } 
\end{eqnarray}
and, therefore, the different moments of the position of the wave packet center are easily calculated to be
\begin{eqnarray} 
\la q(t) \ra &=& q(0) + \dot{q}(0) \frac{1}{\ga} (1- e^{-\ga t} )  
\\
\la \dot{q}(t) \ra &=& \dot{q}(0) e^{-\ga t}  
\\
\la q^2(t) \ra &=& \left(  q(0) + \dot{q}(0) \frac{1}{\ga} (1- e^{-\ga t} ) \right)^2 
- \frac{k_B T}{m \ga^2} ( 3 + e^{-2\ga t} - 4 e^{-\ga t} - 2 \ga t )
\\
\la \dot{q}^2(t) \ra &=& \dot{q}(0)^2 e^{-2\ga t} + \frac{k_B T}{m} ( 1 - e^{-2\ga t})     .
\end{eqnarray}
A second average over all values of $ \dot{q}(0) $ is needed and using the fact that 
\begin{eqnarray} 
\la \dot{q}(0) \ra &=& 0 \label{eq: v0_av} \\
\la \dot{q}(0)^2 \ra &=& \frac{k_B T}{m} \label{eq: v02_av}
\end{eqnarray}
the double averages are then expressed as
\begin{eqnarray} 
\la\la q(t) \ra\ra &=& q(0)   \label{eq: qavav_free}
\\
\la\la \dot{q}(t) \ra\ra &=& 0
\\
\la\la q^2(t) \ra\ra &=& q(0)^2 + 2\frac{k_B T}{m \ga} \left( t - \frac{1-e^{-\ga t}}{\ga} \right) 
\label{eq: q2avav_free}
\\
\la\la \dot{q}^2(t) \ra\ra &=& \frac{k_B T}{m}   .
\end{eqnarray}
For very long times, i.e., times much greater than the relaxation time $ \ga^{-1} $, the mean squared displacement (MSD) becomes 
proportional to time according to
\begin{eqnarray} \label{eq: Eins}
\la\la ( q(t) - q(0))^2 \ra\ra & \approx & 2 D t , \qquad t \gg \frac{1}{\ga}
\end{eqnarray}
where $D$ is known as the diffusion constant which fulfills the so-called Einstein's relation
\begin{eqnarray} \label{eq: diff_cons}
D &=& \frac{k_B T}{m \ga}  .
\end{eqnarray}
From this analysis, one is motivated to define a {\it time-dependent} diffusion coefficient $ D(t) $ as
\begin{eqnarray} \label{eq: diff_cons}
D(t) &=& \frac{1}{2 t} \la\la ( q(t) - q(0))^2 \ra\ra
\end{eqnarray}
whose long time limit gives the diffusion constant. 

The next step is to find the MSD of  Bohmian stochastic trajectories. From Eqs. (\ref{eq: BM_traj}) and (\ref{eq: qavav_free}) we have that
\begin{eqnarray} \label{eq: BM_MSD}
\la\la (x(x^{(0)}, t) - x^{(0)})^2 \ra \ra &=& \la\la  (q(t) - q(0))^2 \ra \ra + 
\left( \frac{\si(t)}{\si(0)} - 1 \right)^2 (x^{(0)} - q(0) )^2    .
\end{eqnarray}
In contrast to the classical result, the Bohmian MSD depends on the initial value of the Bohmian position with respect to the center of the 
wave packet. By taking the last average over initial values of Bohmian positions $ x^{(0)} $ according to the Born distribution and using 
Eq. (\ref{eq: q2avav_free}), the following expression is reached
\begin{eqnarray} \label{eq: BM_MSD_av}
\overline{ \la\la (x(x^{(0)}, t) - x^{(0)})^2 \ra \ra } &=& 
\int dx^{(0)} \rho(x^{(0)}, 0) \la\la (x(x^{(0)}, t) - x^{(0)})^2 \ra \ra 
\\
&=&
2\frac{k_B T}{m \ga^2} ( -1 + e^{-\ga t} + \ga t ) + ( \si(t)- \si(0) )^2    .
\end{eqnarray}
Thus, the time-dependent quantum diffusion coefficient can be written as
\begin{eqnarray} \label{eq: BM_D}
D_{\qm}(t) &=& D_{\cl}(t) + \frac{1}{2 t} ( \si(t) - \si(0) )^2  
\end{eqnarray}
where $D_{\cl} (t)$ can be seen as a time-dependent classical diffusion coefficient. However, according to 
the Pinney equation (\ref{eq: delt}), which does not have analytic solution,  the width behaves in the long time limit as \cite{HaBaSiNa-IJTP-2013}
\begin{eqnarray}
\si(t) & \sim & t ^{1/4}   .
\end{eqnarray}
Thus, at long times, $D(t)$ tends to the classical diffusion constant \cite{Pa-JPA-1973}; that is, the width contribution of the wave packet 
has no influence on the value of the diffusion constant.

It should be noted that an alternative way to find this diffusion constant is by computing the velocity autocorrelation function 
which is expressed in this formalism as \cite{Mi-2018} 
\begin{eqnarray} \label{eq: vel_auto}
\la\la v(x^{(0)}, 0)  v(x^{(0)}, t) \ra \ra &=& \la\la \dot{q}(0)  \dot{q}(t) \ra \ra 
+ \frac{\dot{\si}(t)}{\si(0)} (x^{(0)} - q(0)) \la\la  \dot{q}(0) \ra \ra
= \frac{k_B T}{m} e^{-\ga t} 
\end{eqnarray}
which is just the classical velocity autocorrelation function. Note that this result is independent on $ x^{(0)} $. Hence, averaging over 
Bohmian initial positions does not affect this result. As we mentioned above, one can also compute $D$ via this autocorrelation function
\cite{Tuckerman-book-2010}. In this way, we recover again Einstein's relation
\begin{eqnarray} \label{eq: Diff_const}
D &=& \int_0^{\infty} dt ~ \overline{ \la\la v(x^{(0)}, 0)  v(x^{(0)}, t) \ra \ra  }   
= \frac{k_B T}{m \ga}    .
\end{eqnarray}
%


On the other hand, it is instructive to analyze the position-momentum uncertainty relation in this context. As before, to avoid 
confusion, quantum averages (expectation values) are labelled by bars. From the Gaussian ansatz (\ref{eq: rho_ansatz}), we have that
\begin{eqnarray*}
\overline{\hat{x}} &=& \int dx ~ x \rho(x, t) = q(t)
\\
\overline{\hat{x^2}} &=& \int dx ~ x^2 \rho(x, t) = q(t)^2 + \si(t)^2
\\
\overline{\hat{p}} &=& -i \hb \int dx ~ \psi^*(x, t) \frac{\pa}{\pa x} \psi(x, t)
= \int dx ~ \rho(x, t) \frac{\pa S}{\pa x} = m \dot{q}(t)
\\
\overline{\hat{p^2}} &=& \hb^2 \int dx ~ \big| \frac{\pa \psi(x, t)}{\pa x} \big|^2
= \frac{\hb^2}{4} \int dx ~ \frac{1}{\rho} \left( \frac{\pa \rho}{\pa x} \right)^2
+ \int dx ~ \rho \left(\frac{\pa S}{\pa x}\right)^2 
\\
&=& m^2 \dot{q}(t)^2 + m^2  \dot{\si}(t)^2 + \frac{\hb^2}{ 4 \si(t)^2 }
\end{eqnarray*}
where the hat represents operators. In the last two equations, Eqs. (\ref{eq: BM_vel}) and (\ref{eq: vel}) have been used.
The variances in position and momentum are
\begin{eqnarray}
\la \la (\Delta x)^2 \ra \ra &=& \la \la \overline{\hat{x^2}} \ra \ra  - \la \la \overline{\hat{x}} \ra \ra^2  
= \bigg( \la\la q(t)^2 \ra\ra - \la\la q(t) \ra\ra^2 \bigg) + \si(t)^2
\label{eq: x_var} 
\\
\la \la (\Delta p)^2  \ra \ra &=& \la \la \overline{\hat{p^2}} \ra \ra  - \la \la \overline{\hat{p}} \ra \ra^2  
= m^2 \bigg( \la\la \dot{q}(t)^2 \ra\ra - \la\la \dot{q}(t) \ra\ra^2 \bigg) + m^2  \dot{\si}(t)^2 + \frac{\hb^2}{ 4 \si(t)^2 } 	\label{eq: p_var}  .
\end{eqnarray}
We should have in mind that our quantum system is described by a mixed state and thus some of these variances are intrinsically 
classical \cite{Lu-TMP-2005}. 
%

Now, from Eqs. (\ref{eq: x_var}) and (\ref{eq: p_var}) one can compute the uncertainty product 
$ U(t) = \sqrt{ \la \la (\Delta x)^2 \ra \ra \la \la (\Delta p)^2 \ra \ra } $. From the moments for free particle we have that
\begin{eqnarray} \label{eq: uncer}
U(t)^2 &=& \frac{\hb^2}{4} + m^2 \si(t)^2 \dot{\si}(t)^2 + \left\{ \frac{2}{m \ga^2}(-1 + \ga t + e^{-\ga t}) \left( m^2  \dot{\si}(t)^2 + \frac{\hb^2}{ 4 \si(t)^2 } \right) + m \si(t)^2 \right\} k_B T 
\nonumber \\
&+& \frac{2}{\ga^2}(-1 + \ga t + e^{-\ga t}) ( k_B T )^2  \geq \frac{\hbar^2}{4}  
\end{eqnarray}
%
It is also interesting to mention that in the study of decoherence, Ford and O'Connell \cite{FoCo-JOB-2003} obtained 
\begin{eqnarray} \label{eq: Ford_var}
w^2(t) &=& \overline{  ( \hat{x}(t) - \hat{x}(0) )^2  } + \sigma_0^2 + \frac{ \hb^2 ( 1 - e^{-\ga t} )^2 }{4 m^2 \ga^2 \sigma_0^2}
\end{eqnarray}
for the variance of the probability distribution at time $t$, taking the initial state as a Gaussian wave packet. They provided the analytical 
relation $ \overline { ( \hat{x}(t) - \hat{x}(0) )^2 } = \frac{2 k_B T}{m \ga} \left( t - \frac{1 - e^{-\ga t}}{\ga} \right) $
for the MSD at only high temperatures, $ k_B T \gg \hb \ga $  which is just the classical MSD 
$ \la\la q(t)^2 \ra\ra - \la\la q(t) \ra\ra^2 $
given by second term in the right side of Eq. (\ref{eq: q2avav_free}). In our case, the variance is given by Eq. (\ref{eq: x_var}). They provided an analytical relation for the width given by last terms of Eq. (\ref{eq: Ford_var}). One can obtain such an analytic relation by considering the Caldirola-Kanai approach instead of the Kostin framework for dissipative dynamics. 
%

%
%
%

%
\subsection{Falling particles}  \label{subsec: gr}
%

For a particle in the gravitational potential $ V(x) = m g x $ and from Eq. (\ref{eq: xt_gen_sol}), the solution to the 
corresponding equation of motion is given by
\begin{eqnarray} \label{eq: xt_g_formal_sol} 
q(t) &=& q(0) + \dot{q}(0) \frac{1}{\ga} (1- e^{-\ga t} ) - \frac{g}{\ga^2} ( -1 + \ga t + e^{-\ga t} )
+ \frac{1}{m \ga} \int_0^t d\tau F_r(\tau) e^{-\ga ( t - \tau) } .
\end{eqnarray}
Thus, for the moments we have
\begin{eqnarray} 
\la q(t) \ra &=& q(0) + \dot{q}(0) \frac{1}{\ga} (1- e^{-\ga t} ) - \frac{g}{\ga^2} (-1 + \ga t + e^{-\ga t} )
\\
\la \dot{q}(t) \ra &=& \frac{1}{\ga}( \ga \dot{q}(0) + g - g e^{\ga t} ) e^{-\ga t}  
\\
\la q^2(t) \ra &=& \left(  q(0) + \dot{q}(0) \frac{1}{\ga} (1- e^{-\ga t} - \frac{g}{\ga^2} (-1 + \ga t + e^{-\ga t} ) ) \right)^2 
- \frac{k_B T}{m \ga^2} ( 3 + e^{-2\ga t} - 4 e^{-\ga t} - 2 \ga t )
\\
\la \dot{q}^2(t) \ra &=& \left( \frac{1}{\ga}( \ga \dot{q}(0) + g - g e^{\ga t} ) e^{-\ga t} \right)^2 + \frac{k_B T}{m} ( 1 - e^{-2\ga t})  ,
\end{eqnarray}
and from Eqs. (\ref{eq: v0_av}) and (\ref{eq: v02_av}), one obtains
\begin{eqnarray} 
\la\la q(t) \ra\ra &=& q(0) - \frac{g}{\ga^2} (-1 + \ga t + e^{-\ga t} )   
\\
\la\la \dot{q}(t) \ra\ra &=& \frac{g}{\ga} (-1 + e^{-\ga t} )
\\
\la\la q^2(t) \ra\ra &=& \left( q(0) - \frac{g}{\ga^2} (-1 + \ga t + e^{-\ga t} ) \right)^2 + 2\frac{k_B T}{m \ga^2} ( -1 + e^{-\ga t} + \ga t )
\\
\la\la \dot{q}^2(t) \ra\ra &=&  \frac{g^2}{\ga^2} (1 - e^{-\ga t} )^2 + \frac{k_B T}{m} .
\end{eqnarray}

In the context of the Bohmian stochastic trajectory approach, the mean arrival time at the detector location $x_{\xd}$ is defined by
\begin{eqnarray} \label{eq: mean_ar}
\uptau_{\ar} &=&  
\int_{-\infty}^{\infty} dx^{(0)} \rho(x^{(0)}, 0)  ~ \la \la t_{\ar}(x^{(0)}; \dot{q}(0)) \ra \ra
\end{eqnarray}
where $ t_{\ar}(x^{(0)}; \dot{q}(0)) $ is the arrival time of a particle initially located at the point $ x^{(0)} $ with an initial velocity
$\dot{q}(0)$.  As previously mentioned, it should be remembered that when $ \dot{\si}(0) = 0 $, Bohmian particles move initially with the same 
velocity $\dot{q}(0)$. 

%
\subsection{Parabolic repeller} \label{subsec: PR}
%

The center of the wave packet in the presence of the parabolic repeller $ V(x) = - m \om^2 x^2 /2 $ is obtained from Eq. (\ref{eq: xt_gen_sol}) 
by imposing $ g = 0 $. Then, one has
\begin{eqnarray} 
\la q(t) \ra &=& q(0) \left( \cosh (\Om t) + \frac{\ga}{2 \Om} \sinh (\Om t) \right) e^{-\ga t /2} 
+ \dot{q}(0) \frac{1}{\Om} \sinh (\Om t)  e^{-\ga t /2} 
\label{eq: xt_av_PR}
\\
\la \dot{q}(t) \ra &=& q(0) \frac{\om^2}{\Om} \sinh (\Om t)  e^{-\ga t /2} 
+ \dot{q}(0) \left[ \cosh (\Om t) - \frac{\ga}{2 \Om} \sinh (\Om t) \right] e^{-\ga t /2}
\\
\la q(t)^2 \ra &=& \la q(t) \ra^2
+ \frac{k_B T}{m \om^2} \left[ -1 + e^{-\ga t} \left( 1 + \frac{\ga}{2 \Om} \sinh (2\Om t) + \frac{\ga^2}{2 \Om^2} \sinh^2(\Om t)  \right) \right]
\label{eq: xt2_av_PR}
\\
\la \dot{q}(t)^2 \ra &=& \la \dot{q}(t) \ra^2 +
\frac{k_B T}{m} \left[ 1 + e^{-\ga t} \left( -1 + \frac{\ga}{2 \Om} \sinh (2\Om t) - \frac{\ga^2}{2 \Om^2} \sinh^2(\Om t)  \right) \right] 
\end{eqnarray}
from which by averaging over initial velocities distributed according to the Maxwell-Boltzmann distribution function one reaches
\begin{eqnarray} 
\la \la q(t) \ra \ra &=& q(0) \left( \cosh (\Om t) + \frac{\ga}{2 \Om} \sinh (\Om t) \right) e^{-\ga t /2} 
\label{eq: xt_avav_PR}
\\
\la \la \dot{q}(t) \ra \ra &=& q(0) \frac{\om^2}{\Om} \sinh (\Om t)  e^{-\ga t /2}
\\
\la \la q(t)^2 \ra \ra &=& - \frac{k_B T}{m \om^2} +
\left( q(0)^2 +\frac{k_B T}{m \om^2} \right) \left( \cosh (\Om t) + \frac{\ga}{2 \Om} \sinh (\Om t) \right)^2 e^{-\ga t} 
\label{eq: xt2_avav_PR}
\\
\la \la \dot{q}(t)^2 \ra \ra &=& \frac{k_B T}{m} + \frac{\om^4}{\Om^2} \left( q(0)^2 +
\frac{k_B T}{m \om^2} \right) \sinh^2 (\Om t) ~ e^{-\ga t}    .
\end{eqnarray}
By replacing now $ \om \rightarrow i ~ \om $, the parabolic repeller becomes a simple harmonic oscillator. Thus, by such a replacement, the 
above equations are valid for the over-damped harmonic oscillator which have already been calculated by Chandrasekhar \cite{Ch-RMP-1943} 
in a different way.

Since our noise is Gaussian and white, from the first moments of position, the distribution function $ W_1(q, t; q(0),\dot{q}(0)) $ for $ q $ 
is as follows, 
\begin{eqnarray} \label{eq: W1_func}
W_1(q, t; q(0),\dot{q}(0)) &=&
\frac{1}{\sqrt{2\pi (\la q(t)^2 \ra - \la q(t) \ra^2)} } \exp \left\{ - \frac{ ( q - \la q(t) \ra )^2 }{ 2 (\la q(t)^2 \ra - \la q(t) \ra^2) } \right\}
\nonumber \\
&=& 
\left\{ 2 \pi \frac{k_B T}{m \om^2} \left[ -1 + e^{-\ga t} \left( 1 + \frac{\ga}{2 \Om} \sinh (2\Om t) + \frac{\ga^2}{2 \Om^2} \sinh^2(\Om t)  \right) \right]
\right \}^{-1/2} 
\nonumber \\
& \times & 
\exp \left \{
- \frac{ \left[  q - \left( q(0) \left( \cosh (\Om t) + \frac{\ga}{2 \Om} \sinh (\Om t) \right) e^{-\ga t /2} 
+ \dot{q}(0) \frac{1}{\Om} \sinh (\Om t)  e^{-\ga t /2} \right) \right ]^2 }
{ 2 \frac{k_B T}{m \om^2} \left[ -1 + e^{-\ga t} \left( 1 + \frac{\ga}{2 \Om} \sinh (2\Om t) + \frac{\ga^2}{2 \Om^2} \sinh^2(\Om t)  \right) \right] }
\right \}
\end{eqnarray}
%

\section{Transmission through a parabolic repeller: transmission probability and dwell time} \label{sec: tran_rep}

In this section we are going to study transmission through a parabolic repeller. First for pure dissipative scattering of thermal wave packets and 
then when the fluctuating force is taken into account. In both cases, the transmission probability and dwell time will be expressed by means of
the complementary error function from Bohmian (dissipative and stochastic) trajectories. 

\subsection{Dissipative scattering}

\subsubsection{Transmission probability}

The transmission probability for each element of our ensemble $\psi_{\dot{q}(0), \ga}(x, t)$ is expressed as
\cite{Pa-JPA-1997, BaJa-JPA-1992, Pa-JPA-1990} 
\begin{eqnarray} \label{eq: tran_prob_psi}
P_{\tr}(t; \dot{q}(0), \ga) &=& \frac{ \erf( q_{\ga}(t) / \sqrt{2} \si_{\ga}(t)) - \erf(q(0) / \sqrt{2} \si_0) }{ \erfc(q(0) / \sqrt{2} \si_0)  }
\end{eqnarray}
where $ q_{\ga}(t) $ is given by Eq. (\ref{eq: xbar}) with $ g = 0 $ and $ \si_{\ga}(t) $ is the solution of the generalized Pinney equation 
(\ref{eq: Pinney}). It should be noted that in a transmission process, the incoming wave packet is initially well-localized on the {\it left}, 
$ q(0) < 0 $, of the barrier. If  $\si_0 \ll |q(0)|$, one has
\begin{eqnarray*}
\erf(q(0) / \sqrt{2} \si_0) \approx -1, \qquad \erfc(q(0) / \sqrt{2} \si_0) \approx 2
\end{eqnarray*}
from which  Eq. (\ref{eq: tran_prob_psi}) can be rewritten as
\begin{eqnarray} \label{eq: tran_prob_psi_approx}
P_{\tr}(t; \dot{q}(0), \ga) &\approx & \frac{1}{2}  \erfc \left( -\frac{q_{\ga}(t)}{\sqrt{2} \si_{\ga}(t)} \right)   .
\end{eqnarray}
Now, due to the Maxwell-Boltzmann distribution function (\ref{eq: MB_dis}) for the initial velocities, the time dependent 
thermal transmission probability under the presence of dissipation is given by
\begin{eqnarray} \label{eq: thermal_tran_prob}
P_{\tr}(t; \ga, T_s) &=& \int_{-\infty}^{\infty} d\dot{q}(0) ~ f_{T_s}(\dot{q}(0)) ~ P_{\tr}(t; \dot{q}(0), \ga)
\end{eqnarray}
and from the integral representation of the complementary error function, 
%
%
one can express the corresponding transmission probability as
\begin{eqnarray} 
P_{\tr}(t; \ga, T_s) &=& \frac{1}{\pi} \sqrt{ \frac{m}{2 k_B T_s} } \int_0^{\infty} dy ~ e^{-y^2}
\int_{-\infty}^{\infty} d\dot{q}(0) ~ \exp \left[ - \frac{m\dot{q}(0)^2}{2k_B T_s} - \frac{q_{\ga}(t)^2}{2 \si_{\ga}(t)^2 } + \sqrt{2} \frac{q_{\ga}(t)}{\si_{\ga}(t)} y \right] \nonumber \\
&=& \frac{1}{2} \erfc \left( - \frac{ X_{\ga}(t) }{\sqrt{2} \si_{\ga, T_s}(t)} \right) 
\label{eq: thermal_tran_prob_1}
\end{eqnarray}
where $ X_{\ga}(t) $ is given by Eq. (\ref{eq: Xt_gamma}) with $ g = 0 $ and $ \si_{\ga, T_s}(t) $ by Eq. (\ref{eq: var}).
An alternative way to compute the thermal transmission probability is to make use of $\rho_{T_s}(x, t)$.  
Thus, the thermal transmission probability should be now expressed as
\begin{eqnarray} \label{eq: thermal_tran_prob2}
P_{\tr}(t; \ga, T_s) &=& \int_0^{\infty} dx ~ \rho_{\ga, T_s}(x, t) = 
\frac{1}{2} \erfc \left( - \frac{ X_{\ga}(t) }{\sqrt{2} \si_{\ga, T_s}(t)} \right)    .
\end{eqnarray}
The stationary value of the thermal transmission probability is reached when 
\begin{eqnarray} \label{eq: thermal_tran_prob_st}
P_{\tr}(\ga, T_s) &=& P_{\tr}(t; \ga, T_s) \bigg|_{t\rightarrow \infty}. 
\end{eqnarray}

\subsubsection{Arrival times}

The probability current is used to study the arrival time distribution. In this approach, this distribution is proportional to the modulus of the probability 
current density \cite{artime}. 
Thus, for an ensemble element $ \psi_{\dot{q}(0), \ga}(x, t) $, the arrival time distribution at the detector location $ x_{\xd} $ is given by
\begin{eqnarray} \label{eq: pcd_psi}
\Pi_{\text{A}}(x_{\xd} , t; \dot{q}(0), \ga) &=& \frac{ | j_{\dot{q}(0), \ga}(x_{\xd} , t) | }{ \int_0^{\infty} dt' ~ | j_{\dot{q}(0), \ga}(x_{\xd} , t') | }  
\end{eqnarray}
with the current density being
\begin{eqnarray} \label{eq: current_psi}
j_{\dot{q}(0), \ga}(x , t)  &=&  \frac{\hb}{m} \text{Im} \left\{ \psi^*_{\dot{q}(0), \ga}(x , t) 
\frac{\pa}{\pa x} \psi_{\dot{q}(0), \ga}(x , t)
\right\}
\end{eqnarray}
and therefore for a pure ensemble described by the wave function $ \psi_{\dot{q}(0), \ga}(x, t) $, the mean arrival time at the detector location can be written as  
\begin{eqnarray} \label{eq: mean_ar_psi}
\uptau_{\text{A}}(x_{\xd}; \dot{q}(0), \ga) &=&  \int_0^{\infty} dt' ~ t' ~ \Pi_{\text{A}}(x_{\xd} , t; \dot{q}(0), \ga) .  
\end{eqnarray}
It should be noted that the above expression for the mean arrival time can be uniquely obtained in the framework 
of Bohmian mechanics \cite{Le-PLA-1993,Holland-book-1993} due to the well defined concept of a trajectory.

By averaging now over the Maxwell-Boltzmann distribution, one can calculate thermal mean arrival times when the system is described by 
Eq. (\ref{eq: mat_elem_t}) as follows
\begin{eqnarray} \label{eq: mean_ar_thermal}
\uptau_{\text{A}}(x_{\xd}; \ga, T_s) &=&  \int d\dot{q}(0) ~ f_{T_s}(\dot{q}(0)) ~ \uptau_{\text{A}}(x_{\xd}; \dot{q}(0), \ga) 
\\
&=& \int_0^{\infty} dt' ~ t' \int d\dot{q}(0) ~ f_{T_s}(\dot{q}(0))~ \Pi_{\text{A}}(x_{\xd} , t'; \dot{q}(0), \ga) 
\end{eqnarray}
from which one obtains
\begin{eqnarray} \label{eq: pcd_thermal}
\Pi_{\text{A}}(x_{\xd} , t; \ga, T_s) &=& 
\int d\dot{q}(0) ~ f_{T_s}(\dot{q}(0))~ \Pi_{\text{A}}(x_{\xd} , t; \dot{q}(0), \ga) 
\end{eqnarray}
for the thermal arrival time distribution with dissipation.

\subsubsection{Dissipative dwell time} \label{subsub: tauD}

The question of transit times in a region of space is clear in classical mechanics due to trajectories. The ensemble of incident classical particles  
scattering by a potential barrier is typically split into particles that will transmit through the barrier and particles that will reflect from the barrier. 
Thus, the average dwell time can be separated into two contributions
\begin{eqnarray} \label{eq: class_decom}
\uptau_D &=& P_{\tr} \uptau_{\tr} + P_{\re} \uptau_{\re}
\end{eqnarray}
where $P_{\tr}$ and $P_{\re}$ are the probabilities for transmission and reflection, and $\uptau_{\tr}$ ($\uptau_{\re}$) is the average dwell time for the transmission (reflection) subensembles \cite{Muga-book1}.
The corresponding quantum mechanical dwell time in the space interval $[x_1, x_2]$ for a pure ensemble described by the wave function $ \psi_{\dot{q}(0), \ga}(x, t) $ is
\begin{eqnarray} \label{eq: dwell_time_qm}
\uptau_D(x_1, x_2; \dot{q}(0), \ga) &=& \int_0^{\infty} dt \int_{x_1}^{x_2} dx ~ |\psi_{\dot{q}(0), \ga}(x, t)|^2 ,
\end{eqnarray}
which can be decomposed into reflected and transmitted parts as in Eq. (\ref{eq: class_decom}). It should be noted that this is just one 
of infinitely many possible decompositions \cite{Muga-book1}. 
Eq. (\ref{eq: dwell_time_qm}) can be rewritten as 
\begin{eqnarray} 
\uptau_{\text{D}}(x_1, x_2; \dot{q}(0), \ga) &=& \int_0^{\infty} dt [{\cal Q}(x_1, t; \dot{q}(0), \ga) - {\cal Q}(x_2, t; \dot{q}(0), \ga)] \label{eq: dwell_time_BM}
\end{eqnarray}
where
\begin{eqnarray} 
{\cal Q}(x, t; \dot{q}(0), \ga) &=& \int_x^{\infty} dx'~|\psi_{\dot{q}(0), \ga}(x', t)|^2 = \int_0^t dt'~ j_{\dot{q}(0), \ga}(x, t') 
\label{eq: Q_v0_BM}
\\
&=& \frac{1}{2} \erfc \left( \frac{x - q_{\ga}(t)}{ \sqrt{2} \sigma_{\ga}(t) }  \right)
\label{eq: Q_v0_BM2}
\end{eqnarray}
is the probability of finding the particle beyond the point $x$.
For a mixed ensemble of non-interacting particles, which is initially described by the density operator (\ref{eq: den_mat_t}), the thermal averaging of Eq. (\ref{eq: dwell_time_qm}) over the Maxwell-Boltzmann distribution  yields
\begin{eqnarray} \label{eq: tauD_T1}
\uptau_D(x_1, x_2; \ga, T_s) &=& \int_0^{\infty} dt ~ P_{\ga, T_s}(x_1, x_2, t)
\end{eqnarray}
where
\begin{eqnarray} \label{eq: prob_x1x2}
P_{\ga, T_s}(x_1, x_2, t) &=& \int_{x_1}^{x_2} dx ~ \rho_{\ga, T_s}(x_1, x_2, t)
\end{eqnarray}
gives the probability for finding the particle in the space interval $[x_1, x_2]$ at time $t$ and at the temperature $ T_s $.
By averaging Eq. (\ref{eq: dwell_time_BM}), one has that
\begin{eqnarray} \label{eq: tauD_T2}
\uptau_D(x_1, x_2; \ga, T_s) &=& \int_0^{\infty} dt ~ [ {\cal Q}_{\ga, T_s}(x_1, t) - {\cal Q}_{\ga, T_s}(x_2, t) ]
\end{eqnarray}
an alternative expression for the the dwell time with 
\begin{eqnarray} \label{eq: Q_thermal_def}
{\cal Q}_{\ga, T_s}(x, t) &=& \int_{-\infty}^{\infty} d\dot{q}(0) ~ f_{T_s}(\dot{q}(0)) ~ {\cal Q}(x, t; \dot{q}(0), \ga) .
\end{eqnarray}
By invoking again 
the integral representation of the complementary error function,  Eq. (\ref{eq: Q_thermal_def}) can be rewritten as
\begin{eqnarray} \label{eq: Q_thermal}
{\cal Q}_{\ga, T_s}(x, t) &=& \frac{1}{2} \erfc \left( \frac{ x - q(0) \cosh(\om t) }{\sqrt{2} ~ \si_{\ga, T_s}(t)} \right) .
\end{eqnarray}
One can easily check that the long-time limit of $ {\cal Q}_{\ga, T_s}(x, t) $ for a given $x$ is just the stationary value for the thermal transmission probability 
(\ref{eq: thermal_tran_prob_st}).

It should be noted that at zero temperature, the Maxwell-Boltzmann distribution is just the Dirac delta function centred at 
$ \dot{q}(0) = 0 $, 
\begin{eqnarray} \label{eq: MB_T=0}
f_{T_s}(\dot{q}(0)) \bigg|_{T_s=0} &=& \delta( \dot{q}(0) )
\end{eqnarray}
meaning that instead of a mixed ensemble we have a pure ensemble where all elements of the ensemble are described by the same 
wave function $ \psi_{\dot{q}(0) = 0, \ga}(x, t) $. In this case, thermal quantities are equivalent to those obtained for the pure ensemble with $ \dot{q}(0) = 0 $.

For completeness, the splitting of dwell time into transmission and reflection times is provided in the Appendix 


\subsection{Stochastic dynamics}

Now we consider transmission through the parabolic repeller under the presence of noise. The transmission probability for particles coming 
form the left towards the barrier is just the probability for finding the particle in the right side of the barrier which for the Gaussian wave packet 
(\ref{eq: rho_ansatz}) is known to be
\begin{eqnarray} \label{eq: Tr_prob}
P_{\tr}(t) &=& \int_0^{\infty} dx ~ \rho(x, t) = 
\frac{1}{2} \erfc \left( - \frac{q(t)}{\sqrt{2} \si(t)} \right)
\end{eqnarray}
where we have assumed the incoming Gaussian wave packet is initially {\it well localized} on the left side of the barrier.
Thus, the average transmission probability is obtained from 
\begin{eqnarray} \label{eq: Tr_prob_av}
\la P_{\tr}(t) \ra &=& \frac{1}{2} \int dq ~ W_1(q, t; q(0),\dot{q}(0)) ~ \erfc \left( - \frac{q}{\sqrt{2} \si(t)} \right)
\end{eqnarray}
which from the integral representation of the complementary error function 
can be expressed as
\begin{eqnarray} \label{eq: Tr_prob_av2}
\la P_{\tr}(t) \ra &=& \frac{1}{2} \erfc 
\left( - \frac{ \la q(t) \ra }{\sqrt{2 ( \si(t)^2 + \del_1(t)^2 )} } \right)
\end{eqnarray}
where $ \del_1(t) = \sqrt{ \la q(t)^2 \ra - \la q(t) \ra^2 } $ is the width of $ W_1(q, t; q(0),\dot{q}(0)) $. Now, by averaging over 
initial velocities via the Maxwell-Boltzmann distribution function (\ref{eq: MB_dis})
and using Eqs. (\ref{eq: xt_avav_PR}) and (\ref{eq: xt2_avav_PR}), one arrives at
\begin{eqnarray} \label{eq: Tr_prob_avav}
\la\la P_{\tr}(t) \ra\ra &=& 
\int_{-\infty}^{\infty} d \dot{q}_0 f_T(\dot{q}_0)  \la P_{\tr}(t) \ra
\nonumber \\
&=&
\frac{1}{2} \erfc \left \{ -
\frac{ q(0) \left( \cosh (\Om t) + \frac{\ga}{2 \Om} \sinh (\Om t) \right) e^{-\ga t /2} }
{\sqrt{2}\sqrt{ \si(t)^2 +  \frac{k_B T}{m \om^2} \left[
\left( \cosh (\Om t) + \frac{\ga}{2 \Om} \sinh (\Om t) \right)^2 e^{-\ga t} - 1 \right] } } 
 \right \}
\end{eqnarray}
where for simplicity we have noted the initial velocity as $\dot{q}_0$ instead of $\dot{q}(0)$.
The complementary error function is a decreasing function of its argument. Thus, from Eq. (\ref{eq: Tr_prob_avav}), it is apparent that for a 
given time the transmission probability increases with temperature.

The other interesting quantity is the dwell time which for the Gaussian wave packet (\ref{eq: rho_ansatz}) reduces to
\begin{eqnarray} \label{eq: tauD2}
\uptau_{\text{D}}(x_1, x_2) &=& \int_0^{\infty} dt ~ [{\cal Q}(x_1, t) - {\cal Q}(x_2, t)]
\end{eqnarray}
where
\begin{eqnarray} \label{eq: Q_func}
{\cal Q}(x, t) &=& \frac{1}{2} \erfc \left( \frac{x - q(t)}{ \sqrt{2} \si(t) }  \right)
\end{eqnarray}
is the probability of finding the particle beyond the point $x$. By averaging this function, one has that
\begin{eqnarray} \label{eq: Q_func_av}
\la\la {\cal Q}(x, t) \ra\ra &=& 
\int d \dot{q}_0 f_T(\dot{q}_0) \int dq ~ W_1(q, t; q(0), \dot{q}_0)
\frac{1}{2} \erfc \left(\frac{x-q}{\sqrt{2} \si(t)} \right)
\nonumber \\
&=& \frac{1}{2} \erfc \left( \frac{x - \la\la q(t) \ra\ra}{\sqrt{2 ( \si(t)^2 + \del(t)^2 )} } \right)
, 
\end{eqnarray}
where in the second equality we have used the integral representation 
of the complementary error function and
\begin{eqnarray} \label{eq: width_Temp}
\qquad \del(t)^2 &=& \la\la q(t)^2 \ra\ra - \la\la q(t) \ra\ra^2 .
\end{eqnarray}
Thus, the average value of the dwell time is
\begin{eqnarray} \label{eq: tauD_av}
\la\la \uptau_{\text{D}}(x_1, x_2) \ra\ra &=& \frac{1}{2}
\int_0^{\infty} dt 
\bigg[
 \erfc \big \{ 
\frac{ x_1 - q(0) \left( \cosh (\Om t) + \frac{\ga}{2 \Om} \sinh (\Om t) \right) e^{-\ga t /2} }
{\sqrt{2}\sqrt{ \si(t)^2 +  \frac{k_B T}{m \om^2} \left[
\left( \cosh (\Om t) + \frac{\ga}{2 \Om} \sinh (\Om t) \right)^2 e^{-\ga t} - 1 \right] } } 
 \big \} 
\nonumber \\
& ~ & ~~~~~~~~~~~~ -
\erfc \big\{ 
\frac{ x_2 - q(0) \left( \cosh (\Om t) + \frac{\ga}{2 \Om} \sinh (\Om t) \right) e^{-\ga t /2} }
{\sqrt{2}\sqrt{ \si(t)^2 +  \frac{k_B T}{m \om^2} \left[
\left( \cosh (\Om t) + \frac{\ga}{2 \Om} \sinh (\Om t) \right)^2 e^{-\ga t} - 1 \right] } } 
 \big \}
\bigg]
\end{eqnarray}
%

%
%
\section{Results and discussion} \label{sec: num_cal}
%

For numerical calculations we have used units where $ \hb = m = 1 $. 
The Langevin equation (\ref{eq: xt}) is solved by means of an algorithm proposed in \cite{VaCi-CPL-2006} with initial conditions 
$ q(t)|_{t=0} = q(0) $ and $ \dot{q}(t)|_{t=0} = \dot{q}(0) $. Then, Bohmian stochastic trajectories are computed from Eq.  (\ref{eq: BM_traj}). 
The temperature of the system, $T_s$, and temperature of the environment $ T $ are taken to be the same, otherwise stated. The high temperature
limit is considered when $k_B T >> \hb \ga$.
%

\begin{figure} 
	\centering
	\includegraphics[width=10cm,angle=0]{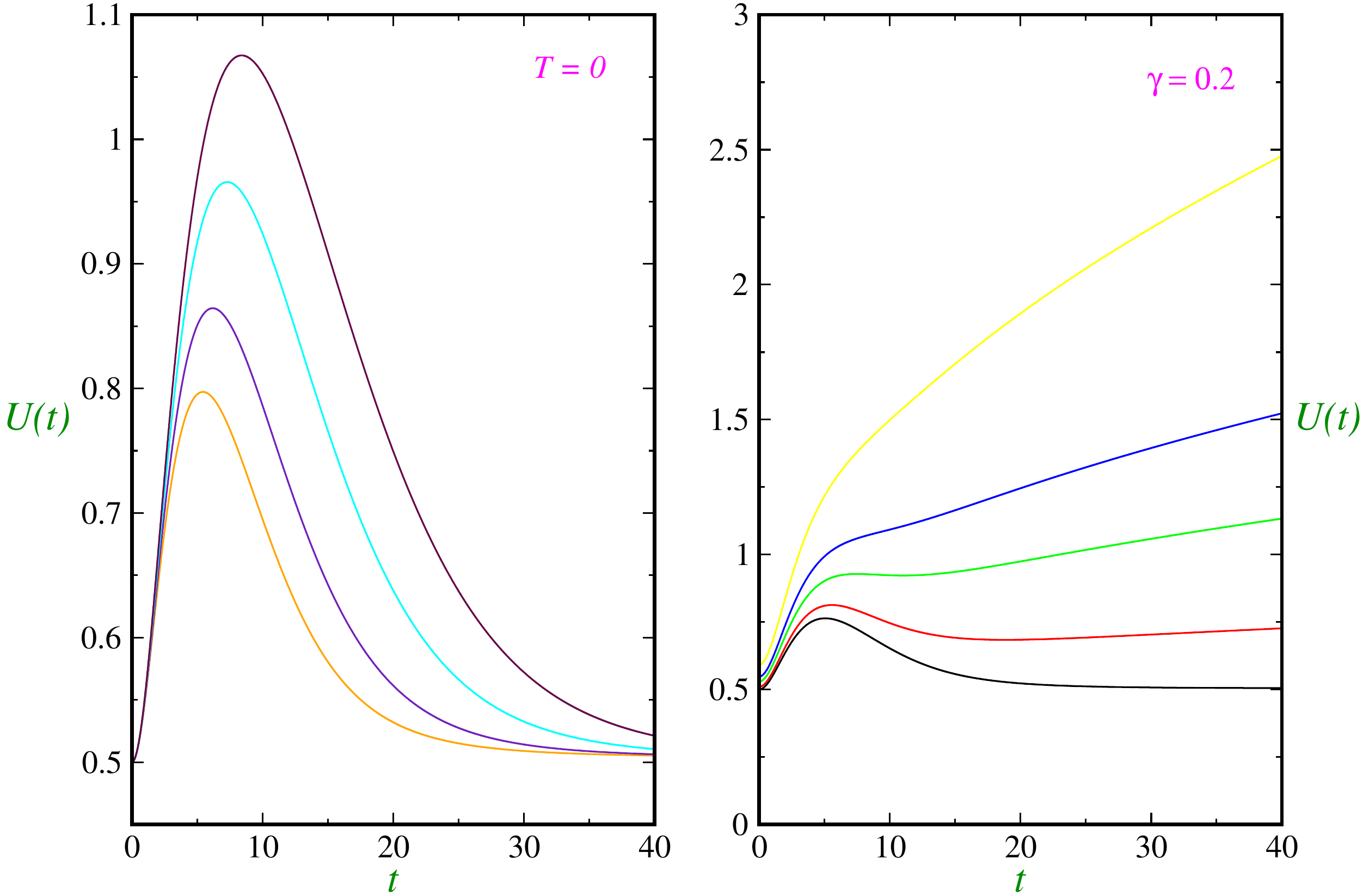}
	\caption{(Color online)
		For the free particle, the uncertainty product $ U(t) = \sqrt{ \la \la (\Delta x)^2 \ra \ra \la \la (\Delta p)^2 \ra \ra } $ versus time is plotted
		for different friction coefficients $ \ga = 0.18 $ (orange curve), $ \ga = 0.15 $ (indigo curve), $ \ga = 0.12 $ (cyan curve) and 
		$ \ga = 0.1 $ (brown curve) at  $ k_B T = 0 $ in the left panel,  and for $\ga = 0.2$ at different temperatures $ k_B T = 0$ (black curve), 
		$ k_B T = 0.01 $ (red curve), $ k_B T = 0.03 $ (green curve), $ k_B T = 0.05 $ (blue curve) and $ k_B T = 0.1 $ (yellow curve) in the right panel. 
	}
	\label{fig: uncertainty} 
\end{figure}

For a free particle, the uncertainty product $ U(t) = \sqrt{ \la \la (\Delta x)^2 \ra \ra \la \la (\Delta p)^2 \ra \ra } $ is plotted versus time in 
Figure \ref{fig: uncertainty} for different values of friction and temperature. In the left panel, the time dependent function $ U(t) $ is shown
for different friction coefficients $ \ga = 0.18 $ (orange curve), $ \ga = 0.15 $ (indigo curve), $ \ga = 0.12 $ (cyan curve) and 
$ \ga = 0.1 $ (brown curve) at  $ k_B T = 0 $. In the right panel, $U(t)$ is plotted for $\ga = 0.2$ 
at different temperatures: $ k_B T = 0$ (black curve), $ k_B T = 0.01 $ (red curve), $ k_B T = 0.03 $ (green curve), $ k_B T = 0.05 $ (blue curve), 
$ k_B T = 0.1 $ (yellow curve). The initial width of the wave packet is chosen to be $ \si_0 = 1 $ and the center is $q(0) =0$. For the pure 
dissipative dynamics (left panel), one clearly sees  that this product increases with time, takes a maximum value and then decreases to reach 
a value slightly greater than $ 0.5 $ asymptotically. This asymptotic value is lower for higher frictions. 
This shows that even in the limit of high frictions corresponding to frequent interactions or observations between the system and the environment, there is no way to reach a minimum value smaller than the limit of 0.5. 
The location of the maximum and its value decreases with friction. 
This should be compared with the non-dissipative free dynamics where the uncertainty product always increases with time. 
In presence of noise (right panel), the corresponding time behaviour is radically different
but the  value of $0.5$ is also preserved.

\begin{figure} 
	\centering
	\includegraphics[width=10cm,angle=0]{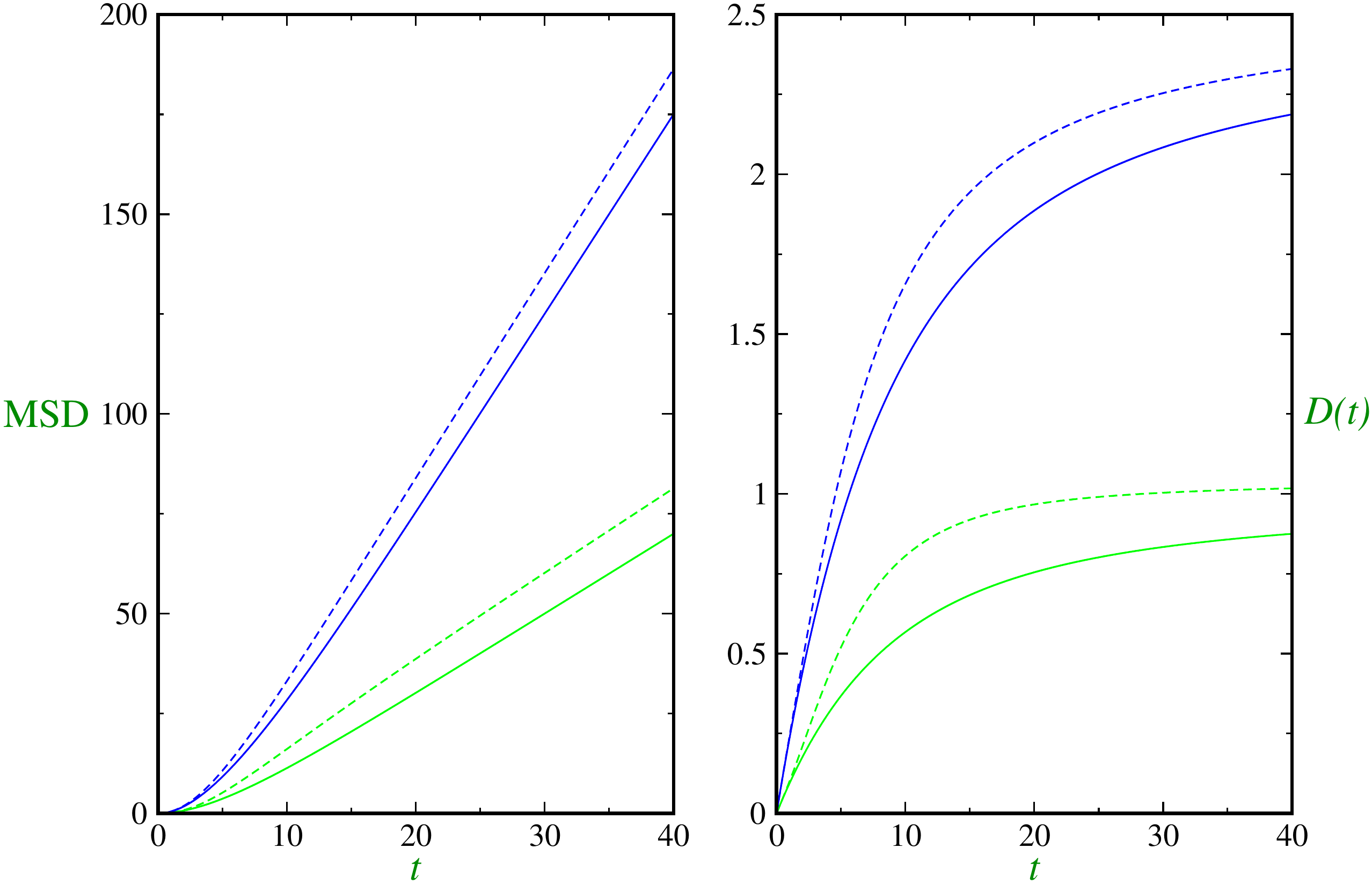}
	\caption{(Color online)
		The time dependent MSD (left panel) and diffusion coefficient (right panel) are plotted for two different temperatures $ k_B T = 0.2 $ 
		(green curves) and $ k_B T = 0.5 $ (blue curves) with a friction of $\ga = 0.2$. The quantum and classical values correspond to the dashed 
        and solid curves, respectively.
	}
	\label{fig: MSD&diff} 
\end{figure}

In Figure \ref{fig: MSD&diff}, the Brownian-Bohmian motion is illustrated. The time dependent MSD (left panel) and diffusion coefficient 
(right panel) are plotted for two different temperatures $ k_B T = 0.2 $ (green curves) and $ k_B T = 0.5 $ (blue curves) with a friction of 
$\ga = 0.2$. The quantum and classical values correspond to the dashed and solid curves, respectively. Parameters of the initial Gaussian wave 
packet are chosen to be $ q(0) = 0 $ and $ \si_0 = 1 $. The time dependent quantum 
diffusion coefficient is always greater than the classical one and both of them reach the same constant value at asymptotic times, the 
diffusion constant. The Einstein relation is fulfilled.

\begin{figure} 
	\centering
	\includegraphics[width=10cm,angle=0]{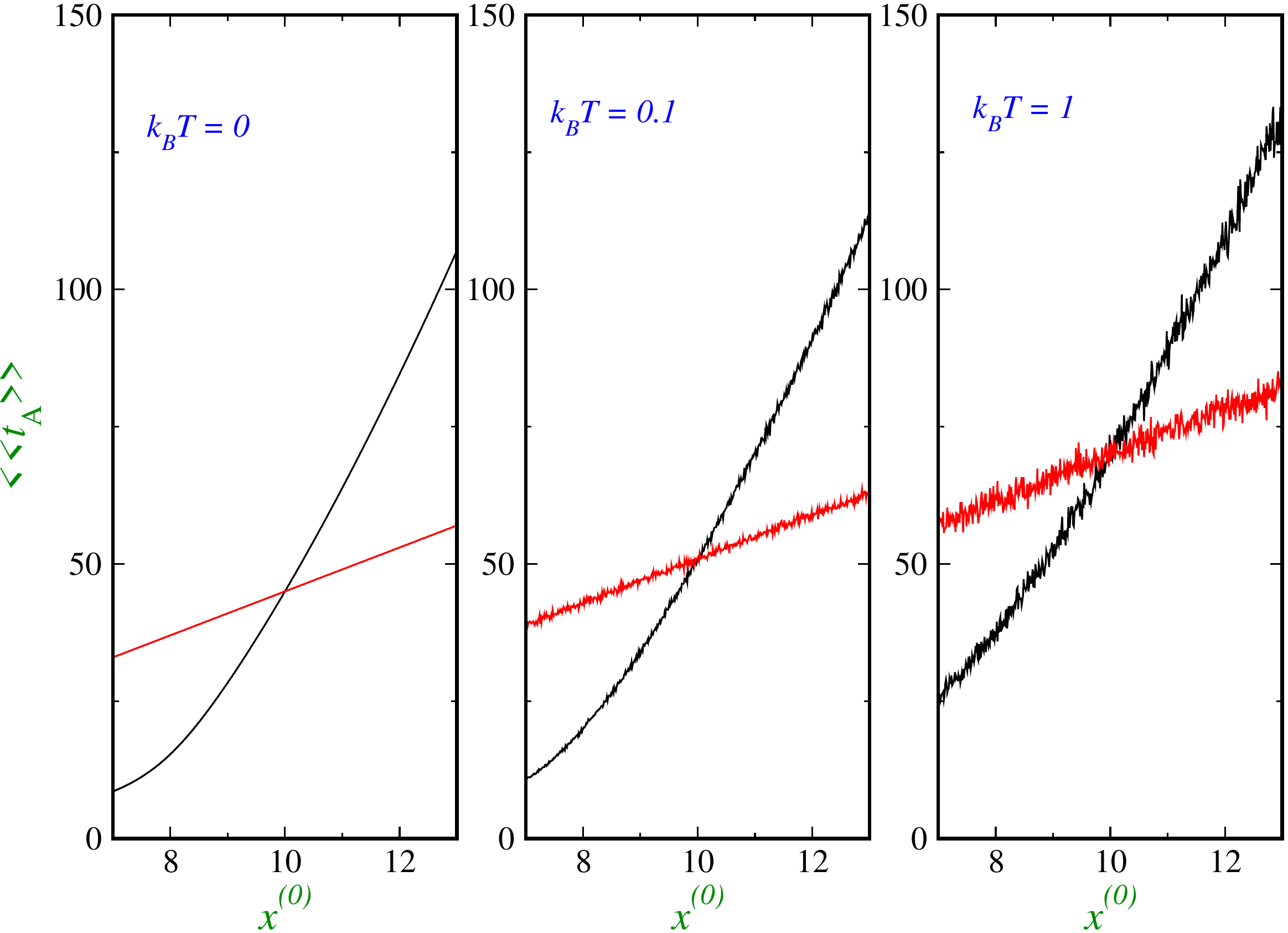}
	\caption{(Color online)
		Thermal arrival times are plotted versus the initial position of the quantum particle for $ \ga =0.2 $ and three different temperatures, 
		$k_B T=0, 0.1,1$. Bohmian and classical results are displayed by black and red curves, respectively.	
	}
	\label{fig: meanar1_y0FF} 
\end{figure}

Let us consider now the falling case. Here, the gravitational strength is considered to be $ g = 0.05 $ and the ground is at $ x_{\xd}=0 $.
In Figure \ref{fig: meanar1_y0FF}, the thermal arrival time is plotted versus the initial position of the quantum particle for $ \ga =0.2 $
and three different temperatures, $ k_B T=0, 0.1,1$.  
The number of Bohmian stochastic trajectories used for averaging is $5,000$. The center of the initial wave packet is located at $ q(0) = 10 $ 
with a width of $ \si_0=1 $. Quantum and classical thermal arrival times are represented by black and red curves, respectively. In our computations, classical arrival times are computed by use of the classical wave function which satisfies the classical Schr\"{o}dinger equation \cite{MoMi-AOP-2018}.
%
From this figure we find that: (1) the thermal arrival time increases with the distance from the ground;
(2) for particles located initially at the center of the packet, the thermal arrival time is the same for both regimes, as should be 
since such particles follows the classical trajectory; (3) particles at the back (front) half of the wave packet move faster (slower) 
in the transition from the quantum regime to the classical one; and (4) the thermal arrival time increases with temperature for a given regime 
and $ x^{(0)} $. The third observation can be explained by the quantum force which takes the form
\begin{eqnarray} \label{eq: qmforce}
F(x^{(0)}, t) &=& \frac{\hb^2}{4 m \si_0 \si(t)^3} ( x^{(0)} - q(0) ) , 
\end{eqnarray}
in the noiseless case. 
From Figure \ref{fig: meanar1_y0FF}, it is possible to explain why the mean arrival time is higher for the quantum regime than for the classical one. 
From Eq. (\ref{eq: mean_ar}) one has
\begin{eqnarray*}
\Delta\uptau_{\ar} &=& \uptau_{\ar, \qm} - \uptau_{\ar, \cl} \approx \int_{-3\si_0}^{3\si_0} dX^{(0)} \rho(X^{(0)}, 0) ~ 
( \la\la t_{\qm}(X^{(0)}) \ra\ra - \la\la t_{\cl}(X^{(0)}) \ra\ra ) 
\\
&=& 
\int_{0}^{3\si_0} dX^{(0)} \rho(X^{(0)}, 0) \{ \Delta t(X^{(0)}) + \Delta t(-X^{(0)}) \}
\end{eqnarray*}
where $ X^{(0)} = x^{(0)} - q(0) $ and $ \Delta t(X^{(0)}) = \la\la t_{\qm}(X^{(0)}) \ra\ra - \la\la t_{\cl}(X^{(0)}) \ra\ra $; and in the second line we have used the symmetry of the initial probability density $ \rho(X^{(0)}, 0) $ around the point $ X^{(0)} = 0 $. But, from this figure 
we have that $ \Delta t(X^{(0)}) > | \Delta t(-X^{(0)}) | $ for $ X^{(0)} > 0  $ and therefore $ \Delta\uptau_{\ar} > 0 $. 
%

\begin{figure} 
	\centering
	\includegraphics[width=10cm,angle=0]{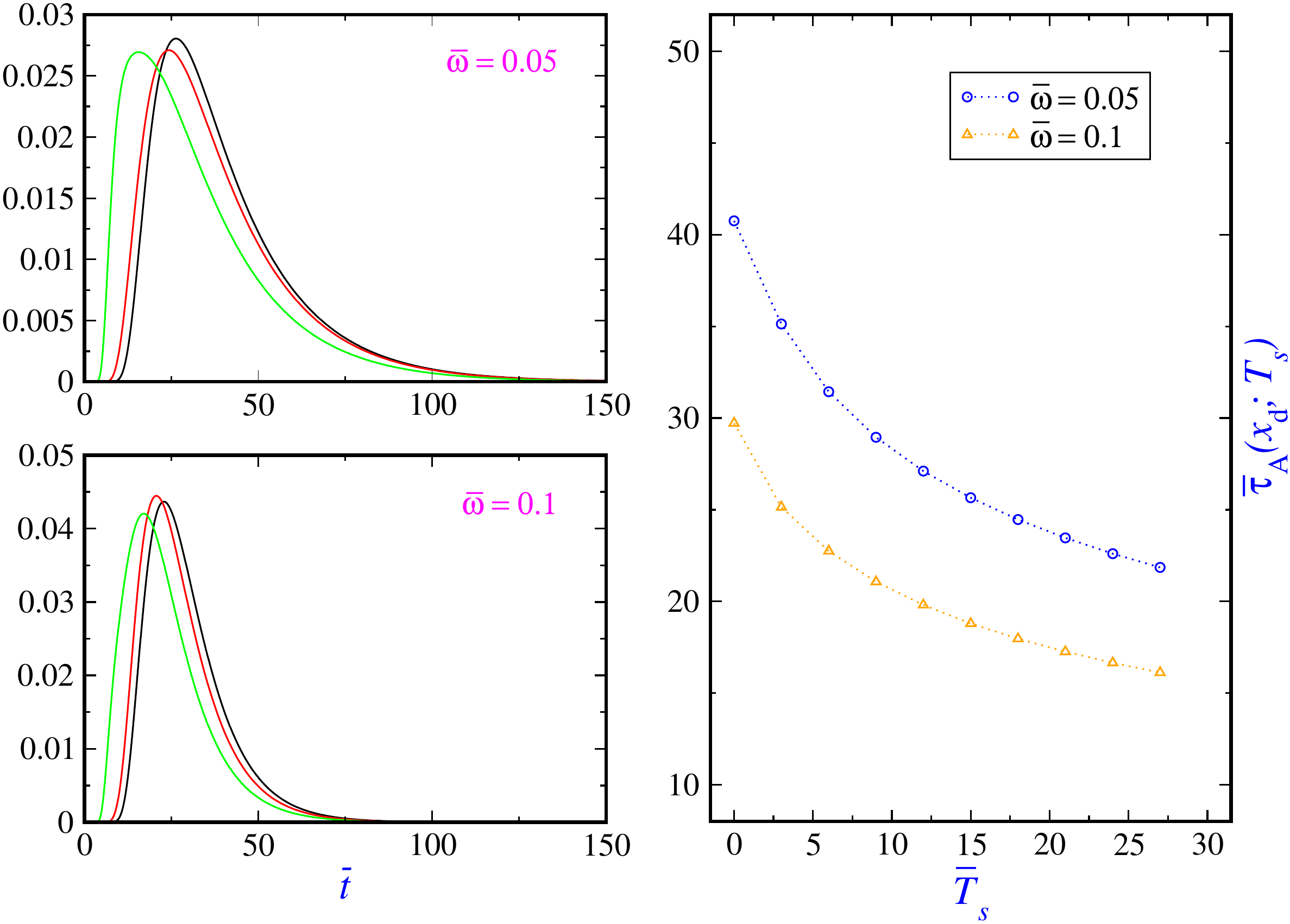}
	\caption{(Color online) 
		Arrival time distributions  for different parabolic repeller strengths, $\bar{ \om } = 0.05$ (left top panel) and $\bar{ \om } = 0.1$ (left 
		bottom panel), for $ \bar{\ga} = 0 $ and $ \bar{T} = 0 $ at  different values of system's temperature: $ \bar{T}_s = 0 $ (black curve), $ \bar{T}_s =  1 $
		(red curve) and $ \bar{T}_s = 5 $ (green curve). The right panel displays mean arrival times at the detector location $\bar{x}_d = 20$ as 
		a function of the system's temperature for the same parabolic repeller strengths.
	}
	\label{fig: arrival} 
\end{figure}
%
%

For computations related to the transmission through the parabolic repeller, dimensionless quantities are used. Thus, we have the following 
reference values: $ \ti{t} = 2 m \si_0^2 / \hb $, $ \ti{\om} = 1 / \ti{t}$ and $ \ti{T} =  \hb^2 / (4m\si_0^2 k_B) $ for times, 
frequencies and temperatures, respectively, their dimensionless units being:  $\bar{\ga}= \ga / \ti{\om} $,  $ \bar{\Om} = \Om / \ti{\om} $, 
$ \bar{T} = T / \ti{T} $. Moreover, lengths are also dimensionless when dividing by $ \si_0 $ 
and denoted by a bar symbol. Scattering particles are electrons and transmission occurs not only via tunnelling. Other parameters  
chosen are: $ \bar{q}(0) = -20 $ for the center of the wave packet,
$ \si_0 = 0.4 ~\AA$ for the initial width, $ \bar{\om} = 0.05 $ and $ \bar{\om} = 0.1 $ for the strengths of the barrier,  and 
$\bar{x}_{\xd} = 20$ for the detector location when computing the arrival times. For computing the thermal characteristic times, the interval 
$ [\bar{x}_1=-1, \bar{x}_2=1] $ is chosen.

In Figure \ref{fig: arrival}, arrival time distributions for $\bar{ \om } = 0.05$ (left top panel) and $\bar{ \om } = 0.1$ 
(left bottom panel) with $ \bar{\ga} = 0 $ and  different values of system's temperature, $ \bar{T}_s = 0 $ (black curve), $ \bar{T}_s =  1 $ (red curve) and 
$ \bar{T}_s = 5 $ (green curve) are plotted for the noiseless case.  In the right panel of the same figure, it is also displayed mean 
arrival times at the detector location as a function of the temperature for the two values of $\bar{\om}$. The maximum of the arrival time 
distribution moves to shorter times as temperature increases. For  a given temperature, this distribution becomes narrower with $ \bar{\om} $. 
As an expected result, the mean arrival time decreases with temperature and the strength of the barrier in this frictionless case.

\begin{figure} 
	\centering
	\includegraphics[width=10cm,angle=0]{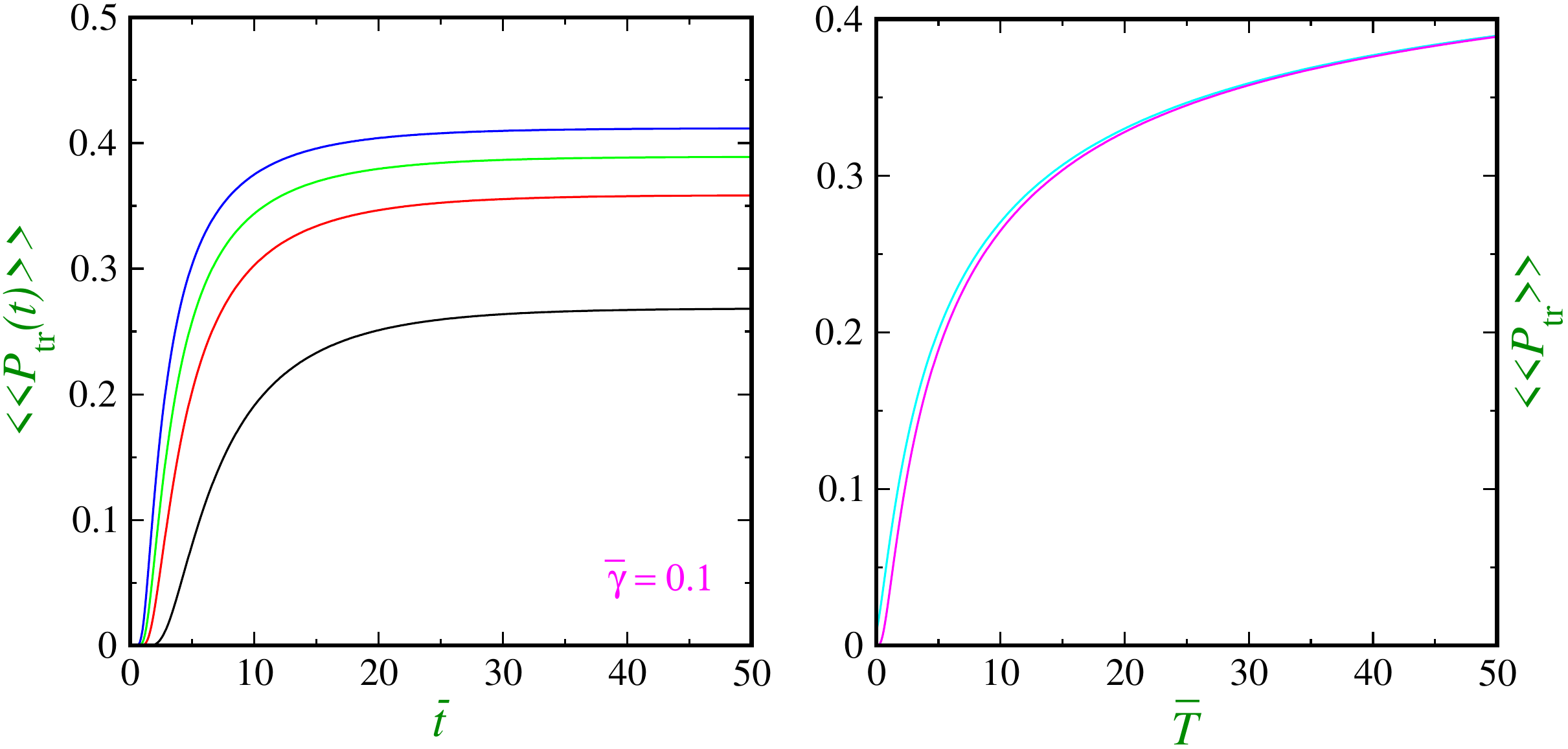} 
	\caption{(Color online) 
		The left panel shows the transmission probability versus dimensionless time for $ \bar{\ga} =0.1 $ at different values of  
		temperature $ \bar{T} =$  $ 10 $ (black curve), $ 30 $ (red curve), $ 50 $ (green curve), $ 80 $ (blue curve). The 
		right panel displays stationary values of the transmission probability versus temperature for two friction coefficients: 
		$\bar{\ga} = 0.05$ (cyan curve) and $\bar{\ga} = 0.5$ (magenta curve).
	}
	\label{fig: Trprob} 
\end{figure}

Transmission probabilities, Eq. (\ref{eq: Tr_prob_avav}), versus dimensionless time have been plotted 
in the left panel of Figure \ref{fig: Trprob} for different values of temperature:  $ \bar{T} =$  $ 10 $ (black curve), $ 30 $ (red curve), 
$ 50 $ (green curve) and $ 80 $ (blue curve) and friction coefficient ${\bar \ga} = 0.1$. The right panel of the same figure
displays time-independent or stationary values of the transmission probability versus temperature for $\bar{\ga} = 0.05$ (cyan curve), 
and $\bar{\ga} = 0.5$ (magenta curve). In both panels, the strength of the parabolic repeller is ${\bar \omega = 0.1}$.
For the initial values chosen for the numerical simulations, part of the transmission process occurs via
tunnelling. Deviation from zero takes place sooner for higher temperatures meaning that particles cross the barrier sooner. 
Transmission increases with temperature for a given friction while it decreases with friction for a given temperature since, as expected, the width of the wave packet decreases with friction, that is, particles become more localized. However, the difference between two different frictions becomes smaller with
temperature. 
As can bee seen from Eq. (\ref{eq: Tr_prob_avav}), the transmission probability reaches $ 0.5 $ at very hight temperatures, 
$ k_B T \gg m \om^2 $, that is, at high temperatures half of the arriving particles cross the barrier's top. Furthermore, for a given friction coefficient
(${\bar \ga} = 0.2$) and temperature ($ \bar{T} =5 $), the transmission probability decreases with the strength of the parabolic repeller, 
going from $0.192$ for ${\bar \omega = 0.1}$  to $2.5 \times 10^{-5}$ for ${\bar \omega = 0.5}$.

\begin{figure} 
	\centering
	\includegraphics[width=8cm,angle=0]{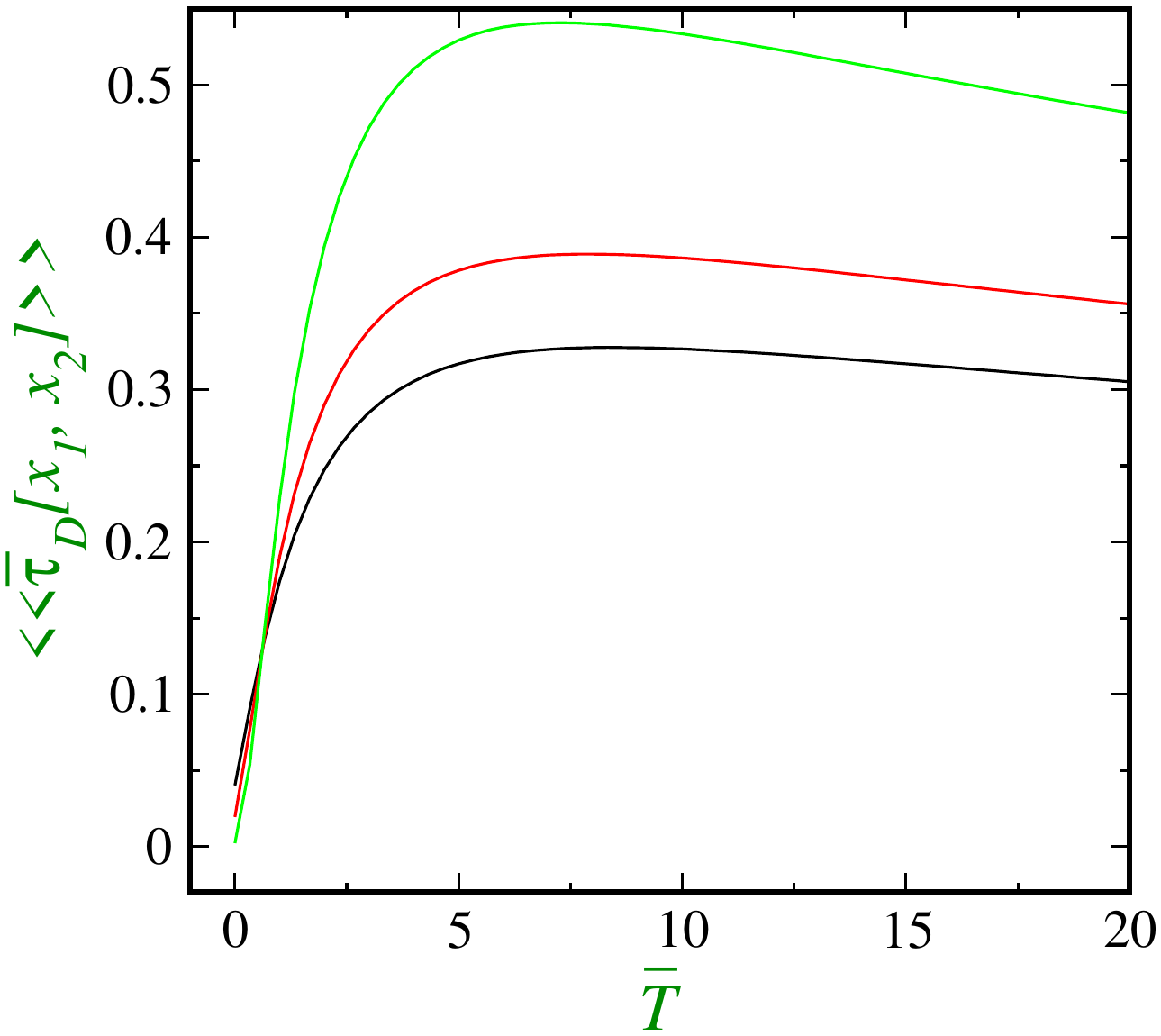} 
	\caption{(Color online) 
		Dwell time in the interval $ [-1, 1] $ versus temperature $\bar T$ for different values of friction: $\bar{\ga} = 0.05$ (black curve), 
		$\bar{\ga} = 0.1$ (red curve) and $\bar{\ga} = 0.2$ (green curve). The strength of the parabolic repeller is ${\bar \omega = 0.1}$ }
	\label{fig: tauD} 
\end{figure}


Dwell times given by Eq. (\ref{eq: tauD_av}) versus temperature $\bar T$ are plotted in Figure \ref{fig: tauD} for a stochastic dynamics.  
with different values of friction: $\bar{\ga} = 0.05$ (black curve),  $\bar{\ga} = 0.1$ (red curve) and $\bar{\ga} = 0.2$ (green curve). 
The strength of the parabolic repeller is ${\bar \omega = 0.1}$ and where the interval used $ [-1, 1] $ has in between the barrier. 
This plot reveals two points. First, at very low temperatures, dwell times decrease with friction while when increasng the temperature, 
this behavior is just the opposite. Second, for a given friction, the dwell time shows a maximum with temperature whose location depends on 
the friction value. In the absence of noise, $T=0$, values of dimensionless dwell time are 0.0402, 0.0194 and 0.002 for 
$\bar{\ga} = 0.05$, $\bar{\ga} = 0.1$ and $\bar{\ga} = 0.2$, respectively. 
As we mentioned at the end of subsection \ref{subsub: tauD}, for $T=0$, we have a pure ensemble where all particles are represented by
motionless wave packets. Thus, the wave packet just spreads over time whose rate decreases with friction as the solution of Pinney equation shows. 
Thus, the probability for finding the particles in the given space interval $[-1, 1]$ for a given time is suppressed by dissipation resulting into 
the reduction of the dwell time. 
By using a different method, such a behavior has already been reported in \cite{KoNiTaHa-PLA-2007} where the mean transversal time 
through a rectangular barrier reduces with dissipation. 
When stochastic fluctuations are taken into account, the width of the probability distribution will have an additional contribution, 
Eq. (\ref{eq: width_Temp}). Friction tends to reduce the width while fluctuations play an opposite role. At low temperatures, the first term is 
the dominant one leading to lower values of dwell times for higher frictions. While at high temperatures, the contribution of thermal fluctuations 
increases and the width dominates (just the opposite role played by the friction) resulting in higher values of dwell time for higher frictions. 
%
In any case, it should be reminded that the three quantities $k_B T/ \hb$, friction $\ga$ and barrier's strength $\om$ are 
important to determine the system dynamics and the behavior of any physical quantity. 

In summary, in this paper after obtaining the SLB equation we have applied it to simple systems where an exact Gaussian wave packet solution exists. 
We have observed that thermal fluctuations only affect the motion of the center of the wave packet while its width remains intact by the noise. 
Application of this equation to other potentials like rectangular barriers with finite extent where the incoming wave packet splits into two parts 
and possible effects due thermal fluctuations on the width of the transmitted packet should be interesting to explore. 

%
%

\vspace{2cm}
\noindent
{\bf Acknowledgement}
\vspace{1cm}

SVM acknowledges support from the University of Qom and SMA support from 
the Ministerio de Ciencia, Innovaci\'on y Universidades (Spain) under the Project 
Project FIS2017-83473-C2-1-P. Valuable discussions about numerical algorithms with M. R. Mozaffari is acknowledged.  

\vspace{2cm}

\appendix

\section{dissipative transmission and reflection times} \label{sec: tran_times}

Due to the non-existence of trajectories in the standard approach to quantum mechanics, the interpretation of Eq. (\ref{eq: dwell_time_qm}) as 
the mean time spent in the region $[x_1, x_2]$ is ambiguous and thus the definition of transit times is not obvious. However, in the context 
of Bohmian mechanics, because quantum trajectories are well defined, a given member of the ensemble of particles can be obviously assigned 
a time of presence. In this Appendix we follow Bohmian mechanics to define transit times in an unambiguous way.

Bohmian dissipative trajectories $ x(x^{(0)}, t; \dot{q}(0), \ga) $ for the Gaussian solution (\ref{eq: rho_ansatz}) are given by 
Eq. (\ref{eq: BM_traj}). The time that a particle, with initial position $ x^{(0)} $, spends in a given space interval $[x_1, x_2]$ can be expressed as
\begin{eqnarray}
t(x_1, x_2; x^{(0)}, \dot{q}(0), \ga) &=& \int_0^{\infty} dt~ \theta( x(x^{(0)}, t; \dot{q}(0), \ga) - x_1 ) ~ \theta( x_2 -x(x^{(0)}, t; \dot{q}(0), \ga) )
\\
&=& \int_0^{\infty} dt \int_{x_1}^{x_2} dx ~ \delta( x(x^{(0)}, t; \dot{q}(0), \ga) - x ) 
\end{eqnarray}
where the $\theta(x)$ is the step function. Then, the mean dwell time is readily calculated as 
\begin{eqnarray} \label{eq: dwell_time}
\uptau_D(x_1, x_2; \dot{q}(0), \ga) &=& \int_{-\infty}^{\infty} dx^{(0)}~ |\psi_{\dot{q}(0), \ga}(x^{(0)}, 0)|^2 t(x_1, x_2; x^{(0)}, \dot{q}(0), \ga)  
\end{eqnarray}
which with respect to the relation $ |\psi_{\dot{q}(0), \ga}(x, t)|^2 = \int dx^{(0)}~ |\psi_{\dot{q}(0), \ga}(x^{(0)}, 0)|^2 \delta( x(x^{(0)}, t; \dot{q}(0), \ga) - x )$ is just the standard relation (\ref{eq: dwell_time_qm}).
For a one-dimensional motion, due to the non-crossing property of Bohmian trajectories, there is always a critical trajectory $ x_c(t; \dot{q}(0), \ga) $ which separates transmitted from reflected trajectories in a scattering problem \cite{SaMi-AOP-2013}. 
Thus, for the {\it stationary} transmission probability one can always write
\begin{eqnarray} \label{eq: tr_prob_BM}
P_{\tr}(\dot{q}(0), \ga) &=& \int_{ x_c(t; \dot{q}(0), \ga) }^{\infty} dx~|\psi_{\dot{q}(0), \ga}(x, t)|^2 .
\end{eqnarray}
By expressing the identity through the Heaviside function,
\begin{eqnarray} 
1 &=& \theta( x - x_c(t; \dot{q}(0), \ga)) + \theta( x_c(t; \dot{q}(0), \ga) - x ) 
\end{eqnarray}
in Eq. (\ref{eq: dwell_time}), the dwell time can be split as  
\begin{eqnarray} \label{eq: dwell_time_BM1}
\uptau_D(x_1, x_2; \dot{q}(0), \ga) &=& P_{\tr}(\dot{q}(0), \ga) ~ \uptau_{\tr}(x_1, x_2; \dot{q}(0), \ga) + P_{\re}(\dot{q}(0), \ga) ~\uptau_{\re}(x_1, x_2; \dot{q}(0), \ga)  
\end{eqnarray}
where the transmission and reflection times are defined respectively as follows
\begin{eqnarray} 
\uptau_{\tr}(x_1, x_2; \dot{q}(0), \ga) &=& \frac{1}{P_{\tr}(\dot{q}(0), \ga)} \int_0^{\infty} dt \int_{x_1}^{x_2} dx ~ 
|\psi_{\dot{q}(0), \ga}(x, t)|^2 ~ \theta( x - x_c(t; \dot{q}(0), \ga)) \label{eq: tr_time_BM1} 
\\
\uptau_{\re}(x_1, x_2; \dot{q}(0), \ga) &=& \frac{1}{P_{\re}(\dot{q}(0), \ga)} \int_0^{\infty} dt \int_{x_1}^{x_2} dx ~ 
|\psi_{\dot{q}(0), \ga}(x, t)|^2 ~ \theta( x_c(t; \dot{q}(0), \ga) - x ) \label{eq: ref_time_BM1}
\end{eqnarray}
in terms of the reflection and transmission probabilities with  $ P_{\re}(\dot{q}(0), \ga) = 1 - P_{\tr}(\dot{q}(0), \ga) $.
These relations show that the calculation of characteristic times within the Bohmian mechanics just requires the knowledge of 
the critical trajectory $ x_c(t; \dot{q}(0), \ga) $. However, it has been proved that there is no need to compute this single trajectory. 
One can always write  \cite{Kr-JPA-2005}
\begin{eqnarray} 
\uptau_{\tr}(x_1, x_2; \dot{q}(0), \ga) &=& \frac{1}{P_{\tr}(\dot{q}(0), \ga)} \int_0^{\infty} dt 
~\left[ \text{min}\{ {\cal Q}(x_1, t; \dot{q}(0), \ga), P_{\tr}(\dot{q}(0), \ga) \} - \text{min}\{ {\cal Q}(x_2, t; \dot{q}(0), \ga), P_{\tr, \ga}(\dot{q}(0), \ga) \} \right]
\nonumber\\
 \label{eq: tr_time_BM} 
\\
\uptau_{\re}(x_1, x_2; \dot{q}(0), \ga) &=& \frac{1}{P_{\re}(\dot{q}(0), \ga)} \int_0^{\infty} dt 
~\left[ \text{max}\{ {\cal Q}(x_1, t; \dot{q}(0), \ga), P_{\tr}(\dot{q}(0), \ga) \} - \text{max}\{ {\cal Q}(x_2, t; \dot{q}(0), \ga), P_{\tr}(\dot{q}(0), \ga) \} \right]
\nonumber\\
 \label{eq: ref_time_BM}
\end{eqnarray}

The thermal transmission and reflection times in the presence of dissipation are respectively given by
\begin{eqnarray}
\uptau_{\tr}(x_1, x_2; \ga, T_s) &=& \int d\dot{q}(0)~ f_{T_s}(\dot{q}(0))~ \uptau_{\tr}(x_1, x_2; \dot{q}(0), \ga)  \label{eq: tau_tr_T}
\\
\uptau_{\re}(x_1, x_2; \ga, T_s) &=& \int d\dot{q}(0)~ f_{T_s}(\dot{q}(0))~ \uptau_{\re}(x_1, x_2; \dot{q}(0), \ga) \label{eq: tau_ref_{T_s}}  .
\end{eqnarray}
%




%
%

\end{document}